\documentclass[a4paper,fleqn]{cas-sc}

\usepackage[authoryear]{natbib}
\usepackage{url,hyperref,lineno}
\usepackage[onehalfspacing]{setspace}
\usepackage{graphicx}

\usepackage{indentfirst}
\usepackage{array,multirow}
\usepackage{times} 
\usepackage{fancyhdr}
\usepackage{booktabs}
\usepackage{multirow}
\usepackage{comment}
\usepackage{siunitx}
\usepackage{multicol}

\usepackage[labelfont=bf,justification=centering]{caption}
\usepackage[labelformat=simple]{subcaption}

\usepackage{algorithmicx,algorithm,algpseudocode}
\usepackage{amsmath,amsfonts,bm}
\usepackage{xcolor}
\usepackage{fontawesome5}

\usepackage[figuresright]{rotating}
\renewcommand{\thefootnote}{\roman{footnote}} 

\usepackage[normalem]{ulem}
\useunder{\uline}{\ul}{}

\newcommand{\ctext}[1]{\raise0.2ex\hbox{\textcircled{\scriptsize{#1}}}}

\newcommand{\eg}{e.\,g.,\ }
\newcommand{\ie}{i.\,e.,\ }

\hypersetup{
    colorlinks=false,
    linkcolor=blue,
    filecolor=magenta,      
    urlcolor=cyan,
    pdftitle={Transportation Research Part F SVC-VV LIU 2023},
    pdfpagemode=FullScreen
    }
    
\def\tsc#1{\csdef{#1}{\textsc{\lowercase{#1}}\xspace}}
\tsc{WGM}
\tsc{QE}
\tsc{EP}
\tsc{PMS}
\tsc{BEC}
\tsc{DE}

\usepackage{float}

\begin{document}
\let\WriteBookmarks\relax
\def\floatpagepagefraction{1}
\def\textpagefraction{.001}

\shorttitle{Motion Sickness Modeling by 6~DoF~SVC-VV Model for using APMV}
\shortauthors{Liu et~al.}


\title [mode = title]{Subjective Vertical Conflict Model with Visual Vertical: Predicting Motion Sickness on Autonomous Personal Mobility Vehicles}

\author[1]{Hailong Liu}[type=editor,
                        auid=000,bioid=1,
                        orcid=0000-0003-2195-3380]
\cormark[1]
\ead{liu.hailong@is.naist.jp}
\author[1]{Shota Inoue}
\author[1]{Takahiro Wada}

\affiliation[1]{organization={Graduate School of Science and Technology, Nara Institute of Science and Technology},
    addressline={8916-5 Takayama-cho}, 
    city={Ikoma},
    state={Nara},
    postcode={630-0192}, 
    country={Japan}}

\cortext[cor1]{Corresponding author}

\begin{abstract}
Passengers of level 3-5 autonomous personal mobility vehicles~(APMV) can perform non-driving tasks, such as reading books and smartphones, while driving.
It has been pointed out that such activities may increase motion sickness, especially when frequently avoiding pedestrians or obstacles in shared spaces.
Many studies have been conducted to build countermeasures, of which various computational motion sickness models have been developed. 
Among them, models based on subjective vertical conflict~(SVC) theory, which describes vertical changes in direction sensed by human sensory organs v.s. those expected by the central nervous system, have been actively developed. 
However, no current computational model can integrate visual vertical information with vestibular sensations. 

We proposed a 6~DoF~SVC-VV model which added a visually perceived vertical block into a conventional 6~DoF~SVC model to predict visual vertical directions from image data simulating the visual input of a human.

In a driving experiment, 27 participants experienced an APMV with two visual conditions: looking ahead (LAD) and working with a tablet device (WAD). 
We verified that passengers got motion sickness while riding the APMV, and the symptom were severer when especially working on it, by simulating the frequent pedestrian avoidance scenarios of the APMV in the experiment.
In addition, the results of the experiment demonstrated that the proposed 6~DoF~SVC-VV model could describe the increased motion sickness experienced when the visual vertical and gravitational acceleration directions were different.
\end{abstract}


\maketitle

\section{INTRODUCTION}

Level 3-5 of driving automation~\citep{SAE_j3016b_2018} are applied to cars and miniaturized personal mobility vehicles~(PMVs)~\citep{morales2017social,liu2020_what_timeing}.
Autonomous PMVs~(APMVs) are expected to be widely used in mixed traffic and shared space conditions, such as sidewalks, shopping centers, stations, and school campuses~\citep{Yoshinori2013, ali2019smart,liu2022implicit}.
Additionally, drivers (passengers) of AVs (including APMVs) are allowed to perform non-driving tasks during autonomous driving~\citep{sivak2015motion,wada2016motion,diels2016self,lihmi}, \eg reading a book~\citep{isu2014quantitative,Sato2022,LIU2022_IV}, watching videos ~\citep{kato2006study,isu2014quantitative}, and playing games~\citep{kuiper2018looking,li2021queasy}.
Unfortunately, the above usage scenarios pose a potential risk of motion sickness for passengers on the APMV by the following reasons:
\begin{itemize}
       \item[1)] In these mixed traffic environments, other traffic participants such as pedestrians, bicycles, and other vehicles, will frequently interact with AVs (including APMVs)~\citep{li2021autonomous,LIU2022_IV}. Passengers may prone to motion sickness owing to the lack of control of the APMV and sensory conflict~\citep{sivak2015motion,wada2016motion,diels2016self,ISKANDER2019716} when the APMV does avoidance behaviors frequently.
       \item[2)] Motion sickness may occur with a high probability when the visual and vestibular systems are stimulated with in-congruent information~\citep{reason1978motion2, diels2016self}.
\end{itemize}

Based on the aforementioned issues, preventing motion sickness can be considered an important challenge for the popularity and widespread use of APMVs.
To address these issues, various computational models have been used to evaluate or estimate the severity of motion sickness.

\subsection{Related works}

The sensory conflict~(SC) theory is widely used to explain the mechanism of motion sickness, which postulates that motion sickness is caused by conflicts between one sensory and expected signals based on previous experience~\citep{reason1978motion2}. 
Oman proposed a mathematical model of the SC theory based on an observer or optimal estimation theoretic framework, in which our motion perception was assumed to be influenced or corrected by the discrepancy between signals from sensory organs and those calculated by internal models in our central nervous systems, and the discrepancy is regarded as a conflict in the SC theory~\citep{oman1990motion}.

Based on SC theory, \citet{bles1998motion} proposed the subjective vertical conflict(SVC) theory, which postulates that motion sickness is caused by conflict between the vertical directions sensed by sensory organs and those estimated by the central nervous systems or their internal models.
Moreover, \citet{bos1998modelling} first proposed the first computational motion sickness model of the SVC theory.
This model simulates the process of motion sickness caused by conflicts between otolith organs(OTO) and their internal models using one-degree-of-freedom~(1~DoF) vertical acceleration inputs.
To express motion sickness caused by multiple degrees-of-freedom of head movement, including head rotation, \citet{kamiji2007modeling} extended the 1~DoF SVC model~\citep{bos1998modelling} to a six-degrees-of-freedom~(6~DoF) SVC model, which included the OTO and semi-circular canals~(SCC) in the vestibular system to accept the three-dimensional(3D) acceleration and 3D angular velocity inputs.
Moreover, based on the \citet{kamiji2007modeling}'s 6~DoF SVC model, \citet{Inoue2022} optimized the structures of conflict feedback integration and parameters to increase the accuracy of the 6~DoF SVC model in presenting the tendency of motion sickness and motion perception of verticality.

Furthermore, \citet{bos2008theory} suggested that visually induced motion sickness can also be explained by SVC theory and proposed a model framework that includes visual information such as visual angular velocities and visual vertical~(VV) information.
However, this study did not consider a concrete method for the application of the experimental data.
As a computational model of motion sickness that can address visual–vestibular interactions, \citet{braccesi2011motion} proposed a motion sickness model based on the interaction between the OTO and visual acceleration.
However, this model does not consider the rotation of the head as perceived by the SCC and visual perception.
To address this issue, \citet{wada2020computational} expanded the original 6~DoF SVC model \citet{kamiji2007modeling} for vestibular motion sickness to include visual-based angular velocity perception using the optical flow method from camera images.

Recalling SVC theory, motion sickness is primarily caused by a conflict of vertical perception between sensor organs and their internal models.
Moreover, some medical studies point to a correlation between the disability in vertical visual perception and motion sickness~\citep{yardley1990motion, michelson2018assessment, guerraz2001visual}.
Therefore, based on \citet{kamiji2007modeling}'s work, we proposed a 6~DoF~SVC-VV model that represents motion sickness owing to the vertical perception from interactions of visual-vestibular systems in our preliminary work~\citep{LIU2022_IV}.
However, in this pre-study, we did not compare the predicted results using the model with the participants' feelings of motion sickness.
Moreover, we only verified the accuracy of the proposed visual vertical prediction method in a wide outdoor environment; its performance in complex indoor environments remains uncertain.

\subsection{Purposes and contributions}

The purposes of this study are as follows. 
\begin{itemize}
\item[1)] Confirming that frequent avoidance behaviors of APMV will cause its passengers getting motion sickness;
\item[2)] Proposing a motion sickness computational model based on the SVC theory which addresses the vertical perception and visual-vestibular interaction.
\end{itemize}

The contributions of this study are as follows. 
\begin{itemize}
\item[1)] We verified that passengers got motion sickness while riding the APMV, particularly working on it, by simulating the frequent pedestrian avoidance scenarios of the APMV in the subject's experiment.
\item[2)] A 6~DoF~SVC-VV model was proposed based on the conventional 6~DoF SVC model~\citep{Inoue2022}, which represents motion sickness owing to the vertical perception from interactions of visual-vestibular systems.
\item[3)] We verified that the proposed 6~DoF~SVC-V model has a performance to represent the increase in motion sickness caused by passengers working with tablet devices while riding in APMV.
\end{itemize}

\section{MOTION SICKNESS MODELING WITH VISUAL VERTICAL ESTIMATION}

\begin{figure}[tb] 
\centering 
\includegraphics[width=1\linewidth]{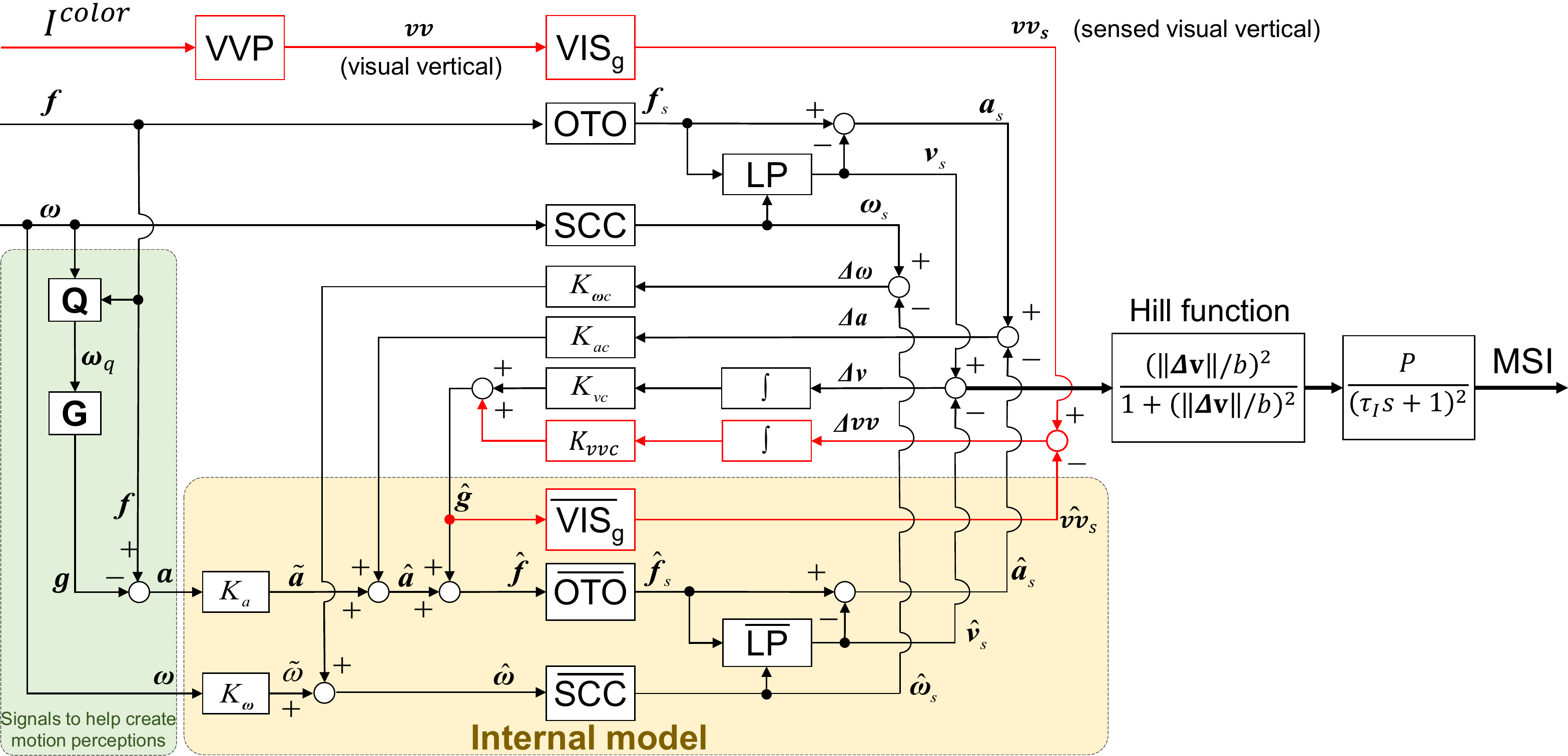}
\caption{Proposed 6~DoF~SVC-VV model: motion sickness computational model considering the vertical sensed using vestibular-visual interactions.} 
\label{fig:model}
\end{figure}

In this study, we propose a 6~DoF SVC-VV model to predict motion sickness incidence(MSI) considering vestibular-visual interaction, as shown in Fig.~\ref{fig:model}.
This model increases the visual vertical~(VV) estimation (shown as red paths) in the 6~DoF SVC model proposed by~\citep{Inoue2022} (shown as black paths).
Moreover, we improve this model to reduce the negative impact of measurement errors in the actual experiment.
In this section, the methods for modeling the visual vertical perception, vestibular system, and interactions with their internal models are presented separately.


\subsection{Visual Vertical Prediction Modeling}

\begin{figure}[htb] 
\centering 
\includegraphics[width=0.9\linewidth]{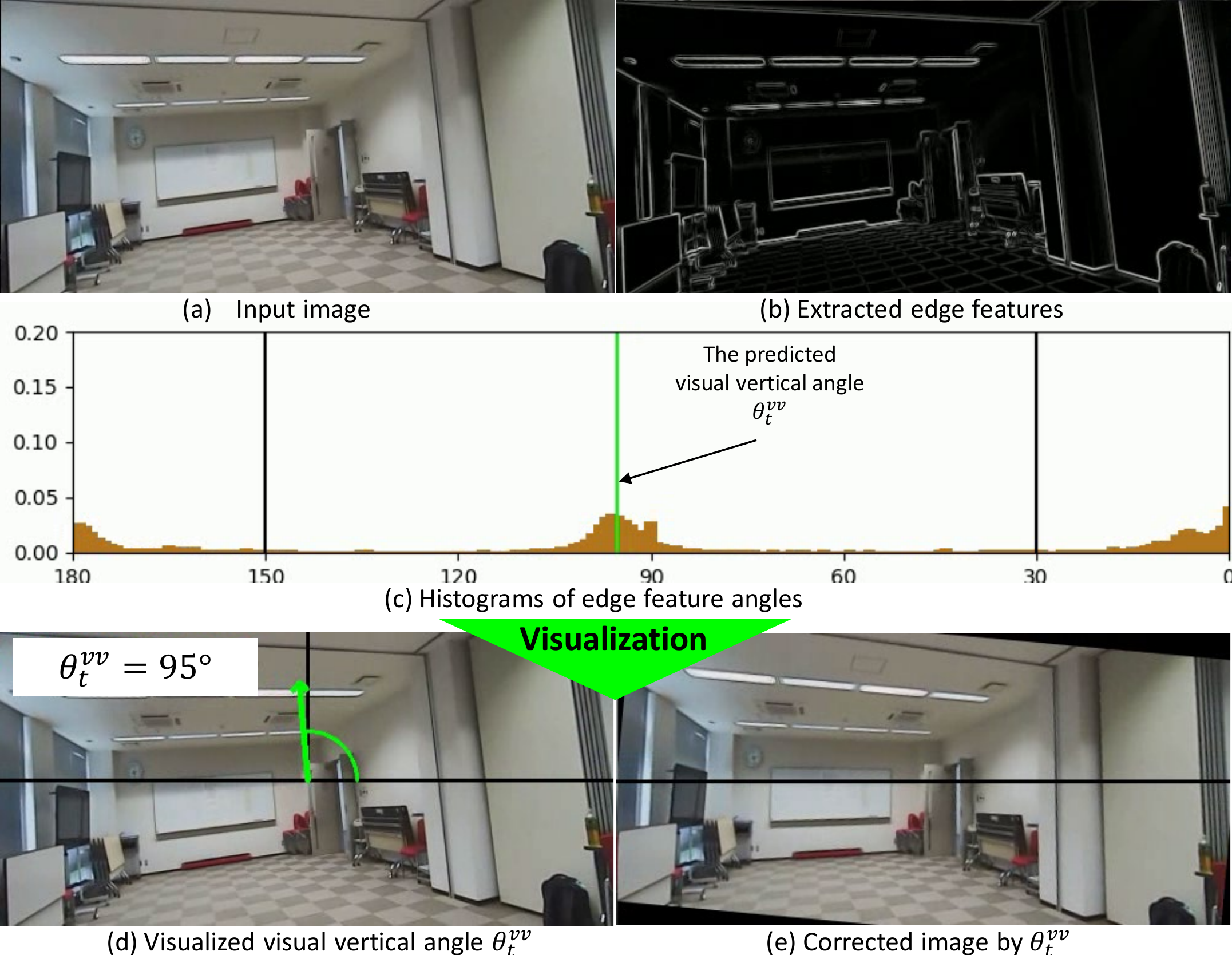}
\caption{An example shows the process of visual vertical prediction in (a)-(c). Visualization results of the predicted visual vertical are shown in (d) and (e).} 
\label{fig:VVP}
\vspace{-2mm}
\end{figure}

Considering that the visual vertical~(VV) is thought to be derived from signals presumed to be parallel or perpendicular to vertical objects, such as buildings or the horizon in the environment~\citep{clark2019mathematical}, a simple image processing method is proposed to estimate the visual vertical by analyzing the directions of the edges of objects in images in our pre-study~\citep{LIU2022_IV}.
In this pre-study, we only considered the usage of APMV in open outdoor areas; thus, the visual vertical direction was calculated based on horizontal edge features such as the horizon and horizontal edge of the building.
However, in indoor scenes with various obstacles such as tables, chairs, and benches, the horizontal edge features change with the viewing angle because of the visual perspective,
particularly in frequent avoidance behaviors of APMV.
Therefore, we take the longitudinal edge features, \eg edges of columns and window frames to calculate the visual vertical direction in this study.
An examples of visual vertical prediction and visualization of the predicted visual vertical is showed in Fig.~\ref{fig:VVP}.
The details of the visual vertical prediction are described as follows.

The \framebox{VVP} block shown in Fig.~\ref{fig:model} represents the process of the visual system predicting the visual vertical from an image.
The proposed visual vertical prediction method is shown in Algorithm~\ref{VV}.
The input is a $\bm{I}_t^{color}\in\mathbb{R}^{H\times W \times3}$ which is defined as a color image in the $t$-th frame captured by a camera attached to the human head to imitate human visual input (see Fig.~\ref{fig:VVP}~(a)).
Then, $\bm{I}_t^{color}$ is preprocessed by converting to a gray-scale image and normalizing through the global maximum and minimum (Algorithm~\ref{VV}, steps 1-2).

Subsequently, Sobel operators are used to compute the gradients in the transverse $\bm{\nabla x}_{t}$ and longitudinal directions $\bm{\nabla y}_{t}$ to detect the edges of the objects in the image (Algorithm~\ref{VV}, steps 3-4).
The gradient magnitudes $M_t$ and angles $\Theta_t$ can be calculated from $\bm{\nabla x}_{t}$ and $\bm{\nabla y}_{t}$ in steps 5-6 of Algorithm~\ref{VV}.
Here, $\odot$ and $\oslash$ are Hadamard products and divisions that are element-wise products and divisions.

\begin{algorithm} [htb]
\setstretch{1.2}
	\caption{Visual vertical direction estimation method for VVP block in Fig.~\ref{fig:model}.} 
	\label{VV} 
\vspace{-3mm}
\begin{multicols}{2}
  {\bf Input:} 
  $\bm{I}^{color}_{t} \in \mathbb{R}^{H \times W \times 3}$ and $\theta^{vv}_0=90$, \\
  \hspace*{9mm} where $H=400, W=1000, t\in\{1,\cdots\, T\}$\\	
  {\bf Output:} $\theta^{vv}$
\begin{algorithmic} [1]
	
\State $\bm{I}^{gray}_{t} \in \mathbb{R}^{H \times W}$ $\gets$ Gray($\bm{I}^{color}$)
	
\State $\bm{I}^{gray}_{t} \in \mathbb{R}^{H \times W}$ $\gets$ Normalization$_{min}^{max}$($\bm{I}^{gray}_{t}$)
	
\State	$\bm{\nabla x}_{t}\in \mathbb{R}^{H \times W}$ $\gets$ Sobel$_x$($\bm{I}^{gray}_{t}$)
  
  \State $\bm{\nabla y}_{t}\in \mathbb{R}^{H \times W}$ $\gets$ $ $Sobel$_y$( $\bm{I}^{gray}_{t}$)
  \State $\bm{M}_{t}=(\bm{\nabla x}_{t}\odot\bm{\nabla x}_{t}+\bm{\nabla y}_{t}\odot\bm{\nabla y}_{t})^{\odot 1/2}$
   \State $\Theta_{t}=  (180 / \pi)\arctan(\bm{\nabla x}_{t}\oslash\bm{\nabla y}_{t})$
  \For {$i = 0$ to $H$} 
    \For {$j = 0$ to $W$}
     
    \State \hspace*{-4mm}$ (\Theta_{t})_{i,j}\gets
        \begin{cases}
       (\Theta_{t})_{i,j} & (0\leq(\Theta_{t})_{i,j}<180)\\
       (\Theta_{t})_{i,j} -180& (180\leq(\Theta_{t})_{i,j}<360)\\
        0 & ((\Theta_{t})_{i,j}=360)
       \end{cases}$
    \EndFor
 \EndFor

  \State $\bm{M}_{t}$ $\gets$ Normalization$_{min}^{max}$($\bm{M}_{t}$)
  
  
  \For {$d = 0$ to 179}
    \State$(\bm{\theta}^{hist}_{t})_d \gets \sum_{i=0}^{H}\sum_{j=0}^{W}\bm{1}_d[(\Theta_{t})_{i,j}] (\bm{M}_{t})_{i,j}$,
    \Statex \hspace*{4mm} where $\bm{\theta}^{hist}_{t}\in\mathbb{N}^{180}$
  \EndFor
  
  \State $\bm{c}^{sort}_{t}\in \mathbb{N}^{121}\gets$Sort($(\bm{\theta}^{hist}_{t})_{29:149}$)
  
  \State $\bm{\theta}^{sort}_{t}\in \mathbb{N}^{121}\gets$ argSort$((\bm{\theta}^{hist}_{t})_{29:149})$ 
  
  \State $\bm{c}^{best3}_{t}\in \mathbb{R}^{3}\gets(\bm{c}^{sort}_{t})_{119:121}/\sum_{i=119}^{121}(\bm{c}^{sort}_{t})_i$ 

  \State $\bm{\theta}^{best3}_{t}\in \mathbb{N}^{3}\gets(\bm{\theta}^{sort}_{t})_{119:121}$ 
  \State $\theta^{vv}_t\gets \bm{\theta}^{best3}_{t} \cdot \bm{c}^{best3}_{t}+30$
  
\State $\theta^{vv}_t\gets
        \begin{cases}
       0.7~ \theta^{vv}_t+0.3~\theta^{vv}_{t-1} & (|\theta^{vv}_t-\theta^{vv}_{t-1}|\leq \ang{4})\\
       0.2~ \theta^{vv}_t+0.8~\theta^{vv}_{t-1} & (|\theta^{vv}_t-\theta^{vv}_{t-1}|>\ang{4})\\
       \end{cases}$

  
  \State $\bm{vv}_t =\begin{bmatrix} vv_t^x\\vv_t^y\\vv_t^z\end{bmatrix}
\gets\begin{bmatrix} 9.81~\cos(\theta^{vv}_t~\pi / 180)\\
9.81~\sin(\theta^{vv}_t~\pi / 180)\\
0\end{bmatrix}$

   \State $\bm{vv}(t)\gets$ZOH$(\bm{vv}_t)$
   \end{algorithmic} 
   \end{multicols}
   \vspace{-3mm}
\end{algorithm}

\clearpage

In $\Theta_t$, we equate [\ang{360}, \ang{180}] to [\ang{0}, \ang{179}], because the angle of a person's neck usually does not exceed \ang{180} (Algorithm~\ref{VV}, steps 8-11).
Further, as the magnitude of the gradient is larger, the edge becomes more likely (see Fig.~\ref{fig:VVP}~(b)).
The gradient magnitudes $\bm{M}_{t}$ are normalized through the global maximum and minimum to $[0,1]$ (Algorithm~\ref{VV}, step 12).

Next, as shown in Fig.~\ref{fig:VVP}~(c), the histogram of gradient angles is calculated using an indicator function from $\Theta_t$ with its weight matrix $\bm{M}_{t}$ (Algorithm~\ref{VV}, steps 13-15).
The number of bins in the histogram is set to 180.
After calculating the histogram, the gradient's angles in the range [\ang{30}, \ang{150}] are sorted in ascending order by their counts (Algorithm~\ref{VV}, steps 16-17).
We assume that the passenger head will not rotate out of the range [\ang{30}, \ang{150}] in most driving situations.

Subsequently, the best three angles~$\bm{\theta}^{best3}_{t}$ are selected based on the highest three counts~$\bm{c}^{best3}_{t}$ (Algorithm~\ref{VV}, steps 18-19).
Then, the direction of the visual vertical $\theta^{vv}_t$ is calculated as Step 20 of Algorithm~\ref{VV}, in which $\bm{c}^{best3}_{t}$ can be considered as the weight of $\bm{\theta}^{best3}_{t}$.

Meanwhile, the direction of the visual vertical $\theta^{vv}_t$ is also affected by the direction of the visual vertical in the previous frame, that is, $\theta^{vv}_{t-1}$.
As shown in Step 21 of Algorithm~\ref{VV}, we choose two different strategies to update $\theta^{vv}_t$ to reduce the instability due to the prediction errors.
An example to show the visualization of $\theta^{vv}_t$ in Fig.~\ref{fig:VVP}~(d) and (e).

In Step 23 of Algorithm~\ref{VV}, the visual vertical vector $\bm{vv}=[vv_x, vv_y, vv_z]^T$ is calculated from $\theta^{vv}_t$ with a fixed L2 norm $9.81~m/s^2$.
Notably, the value on the z-axis of $\bm{vv}$, \ie $vv_z$, should be $0$ because $\theta^{vv}$ is the rotation angle on the x-y plane of the head coordinate system.

In Step 24 of Algorithm~\ref{VV}, as the 6~DoF SVC model is a continuous-time system, $\bm{vv}_t$ is a discrete variable estimated from an image, 
a zero-order holder (ZOH) is used to convert $\bm{vv}_t$ into a continuous variable $\bm{vv}(t)$.

After block \framebox{VVP}, block \framebox{{\textbf{VIS$_g$}}} transfers $\bm{vv}$ to the sensed visual vertical $\bm{vv}_s$.
Note that the vertical signals $\bm{vv}_s$ and $\bm{v}_s$ sensed by the visual and vestibular systems are assumed to be the 3D.
For simplicity, a $3 \times 3$ identity matrix is used as the transform matrix $\bm{T}_{vis}$ in this study.
Therefore, 
\begin{eqnarray}
\bm{vv}_s &=& \bm{T}_{vis}~\bm{vv} \nonumber\\
&=&	\begin{bmatrix}      1 & 0 &0\\      0 & 1 & 0\\      0 & 0 &1      \end{bmatrix}  \begin{bmatrix} vv_x\\ vv_y\\ vv_z\end{bmatrix}.
\end{eqnarray}

\subsection{Vestibular System Modeling}

The vestibular system is mainly composed of otolith organs and semi-circular canals.
As shown in Fig.~\ref{fig:model}, the otolith organ is modeled as a \framebox{\textbf{OTO}} block.
Its input is the gravity-inertial acceleration (GIA) in 3 DoF, which is $\bm{f=a+g}$.
Here, $\bm{a}$ is the inertial acceleration, and $\bm{g}$ is the gravitational acceleration (upward).
Refer to \citep{kamiji2007modeling}, in the \framebox{\textbf{OTO}} block, a $3 \times 3$ identity matrix is used for transforming $\bm{f}$ to the sensed GIA $\bm{f}_s$ as
\begin{equation}
\bm{f}_s=\begin{bmatrix}      1 & 0 &0\\      0 & 1 & 0\\      0 & 0 &1      \end{bmatrix}\bm{f}. 
\end{equation}

Further, the \framebox{\textbf{SCC}} block contained semi-circular canals.
It receives angular velocity $\bm{\omega}$ in 3 DoF and transforms it into the sensed angular velocity $\bm{\omega}_s$ using a transfer function~\citep{merfeld1995modeling}:
\begin{equation}
\bm{\omega}_s=\frac{\tau_a\tau_d s^2}{(\tau_a s+1)(\tau_d s+1)}\bm{\omega}.
\label{Eq:SCC}
\end{equation}


Subsequently, block \framebox{\textbf{LP}} represents the otolith-canal interaction that estimates the detected vertical signal $\bm{v}_s$ from $\bm{f}_s$ and $\bm{\omega}_s$ by updating the raw~\citep{bos2002}:
\begin{equation}
\frac{d\bm{v}_s}{dt}=\frac{1}{\tau}(\bm{f}_s-\bm{v}_s)-\bm{\omega}_s \times \bm{v}_s.
\end{equation}

Moreover, the sensed inertial acceleration $\bm{a}_s$ can be calculated as 
\begin{equation}
\bm{a}_s=\bm{f}_s-\bm{v}_s.
\end{equation}

\subsection{Internal models}

The internal model is a hypothetical central neural representation of anticipatory information generated by the central nervous system concerning the sensory organs.
As shown in Fig.~\ref{fig:model}, the 6 DoF SVC-VV model has three internal models.
Specifically, the internal models of \framebox{\textbf{SCC}}, \framebox{\textbf{OTO}}, and \framebox{{\textbf{VIS$_g$}}} are modeled as blocks of \framebox{$\overline{\textbf{SCC}}$}, \framebox{$\overline{\textbf{OTO}}$}, and  \framebox{$\overline{\bf{VIS_g}}$}, respectively.
Furthermore, the low-path filter in the vestibular system, which separates the perceived signals of vertical and linear acceleration, is also modeled as \framebox{$\overline{\textbf{LP}}$} in the internal model.

For blocks \framebox{$\overline{\textbf{SCC}}$} and \framebox{$\overline{\textbf{OTO}}$},
there are two types of inputs.
One type includes a variety of signals to help create motion perceptions, such as motion predictions~\citep{wada2021computational} and efference copy~\citep{jeannerod2006motor}, which is an internal copy of the neural signal that generates the movement. 
The other types are obtained from the feedback of conflicts between the sensor organs and their internal model.

The \framebox{$\overline{\textbf{SCC}}$} represents the internal model of SCC, which transforms angular velocity $\hat{\bm{\omega}}$ predicted through the internal model to the sensed angular velocity $\hat{\bm{\omega}}_s$ using a transfer function~\citep{merfeld1995modeling}:
\begin{eqnarray}
\hat{\bm{\omega}}_s=\frac{\tau_ds}{\tau_ds+1}\hat{\bm{\omega}}.
 \end{eqnarray}
The predicted $\hat{\bm{\omega}}$ combines the angular velocity signal $\bm{\Tilde{\omega}}$ and the feedback of the difference $\mathit{\Delta}\bm{\omega}$ between sensory information ${\bm{\omega}}_s$ and estimated information $\hat{\bm{\omega}}_s$, that is, $\mathit{\Delta}\bm{\omega}=\bm{\omega_s}-\bm{\hat{\omega}_s}$. 
To simplify the generation process of a variety of signals to help create motion perceptions, this study uses the angular velocity of the head $\bm{\omega}$ as input. 
Therefore,
\begin{eqnarray}
\hat{\bm{\omega}}
&=&\bm{\Tilde{\omega}} +K_{\omega c}\mathit{\Delta}\bm{\omega} \nonumber \\
&=& K_{\omega}\bm{\omega}+K_{\omega c}\mathit{\Delta}\bm{\omega} \nonumber \\
&=& K_{\omega}\bm{\omega}+K_{\omega c}(\bm{\omega_s}-\bm{\hat{\omega}_s}).
\end{eqnarray}

The \framebox{$\overline{\textbf{OTO}}$} represents the internal model of OTO, which transforms predicted GIA $\hat{\bm{f}}$ into the expected afferent signal of GIA $\hat{\bm{f}}_s$ using a $3 \times 3$ identity matrix, that is, 
\begin{eqnarray}
\hat{\bm{f}}_s=\begin{bmatrix}      1 & 0 &0\\      0 & 1 & 0\\      0 & 0 &1      \end{bmatrix}\hat{\bm{f}}.
\end{eqnarray}
The predicted $\hat{\bm{f}}$ using the internal model can be calculated as follows:
\begin{eqnarray}
\hat{\bm{f}}=\hat{\bm{g}}+\hat{\bm{a}},
\label{Eq:hat_f}
\end{eqnarray}
where $\hat{\bm{a}}$ and $\hat{\bm{g}}$ are the gravitational and linear accelerations predicted by the internal model.
Specifically,
\begin{eqnarray}
\hat{\bm{a}}&=&\Tilde{\bm{a}}+K_{ac}\mathit{\Delta}\bm{a} \nonumber \\
&=& K_a\bm{a}+K_{ac}\mathit{\Delta}\bm{a}\nonumber \\
&=& K_a\bm{a}+K_{ac}(\bm{a_s}-\bm{\hat{a}_s});
\label{Eq:hat_a}
\end{eqnarray}
In part, the acceleration ${\bm{a}}$ is calculated from measured GIA $\bm{f}$, that is, ${\bm{a}}=\bm{f}-\bm{g}$.
According to~\citep{kamiji2007modeling}, gravitational acceleration $\bm{g}$ is calculated from $\bm{\omega_q}$ using the following update law:
\begin{eqnarray}
\frac{d\bm{g}}{dt}=-\bm{\omega} \times \bm{g}.
\end{eqnarray}
However, in practice, noise exists in $\bm{\omega}$ obtained from the IMU.
This results in a drift in $\bm{g}$ because the noise in $\bm{\omega}$ is also integrated.

To solve this problem, $\bm{\omega_q}$ is a calibrated angle velocity obtained using a complementary filter~\citep{Wetzstein2017VirtualRC} to reduce the sensing tilt from the IMU.
Complementary filter $Q(\bm{\omega}, \bm{f})$ outputs a quaternion vector $\bm{q} \in \mathbb{R}^4$ to present the orientation calculated from the $\bm{\omega}$ and $\bm{f}$:
\begin{eqnarray}
 \bm{q}=Q(\bm{\omega}, \bm{f}).
 \label{Eq:quaternion}
\end{eqnarray}
Then, $\bm{\omega_q}$ can be approximated as
\begin{eqnarray}
\bm{\omega_q} =\begin{bmatrix}     0 & 1 & 0 &0\\   0 &   0 & 1 & 0\\    0 &  0 & 0 &1      \end{bmatrix} (2\frac{d\bm{q}}{dt}\circ\bm{q}^{-1}),
 \label{Eq:omega_q}
\end{eqnarray}
where $\circ$ denotes the quaternion product.
Thus, the new update law for $\bm{g}$ is
\begin{eqnarray}
 \frac{d\bm{g}}{dt}=-\bm{\omega_q} \times \bm{g}.
 \label{Eq:g_new}
\end{eqnarray}
Then, $\bm{g}$ is normalized in a fixed L2 norm $9.81~m/s^2$ by
\begin{eqnarray}
\bm{g}=9.81~\frac{\bm{g}}{||\bm{g}||}.
\end{eqnarray}

In addition, the $\hat{\bm{g}}$ in part of $\hat{\bm{f}}=\hat{\bm{g}}+\hat{\bm{a}}$ (\ie Eq.~\ref{Eq:hat_f}) is calculated via
\begin{eqnarray}
\bm{\hat{g}}&=&\bm{K}_{vvc} \int_{0}^{t} \mathit{\Delta} \bm{vv} dt+\bm{K}_{vc} \int_{0}^{t} \mathit{\Delta} \bm{v} dt \nonumber \\
&=& \bm{K}_{vvc} \int_{0}^{t} (\bm{vv}_{s}-\bm{\hat{vv}}_{s} ) dt+\bm{K}_{vc} \int_{0}^{t} (\bm{v}_s-\bm{\hat{v}}_s) dt.
\label{Eq:hat_g}
\end{eqnarray}
Here, $\mathit{\Delta} \bm{v}$ and $\mathit{\Delta} \bm{vv}$ are the conflicts of the vertical and visual vertical signals between the sensor organs and their internal model, respectively. 
The $\hat{\bm{v}}_s$ is calculated by \framebox{$\overline{\textbf{LP}}$} block by following update law~\citep{bos2002}:
\begin{eqnarray}
\frac{d\hat{\bm{v}}_s}{dt}=\frac{1}{\tau}(\hat{\bm{f}}_s-\hat{\bm{v}}_s)-\hat{\bm{\omega}}_s \times \hat{\bm{v}}_s,
\end{eqnarray}
which is the same as the update law for $\bm{v}_s$.
Subsequently, $\hat{\bm{a}}_s$ in Eq. ~\ref{Eq:hat_a} is calculated using $\hat{\bm{a}}_s=\hat{\bm{f}}_s-\hat{\bm{v}}_s$.

Meanwhile, $\bm{\hat{vv}}_{s} $ represents the sensed visual vertical in the internal model, which is calculated by block \framebox{$\overline{\bf{VIS_g}}$} using the vertical sensed although visual-vestibular interaction, that is, $\hat{\bm{g}}$.
Thus,
\begin{eqnarray}
\hat{\bm{vv}}_s = \bm{T}_{\overline{vis}}~\hat{\bm{g}}=	\begin{bmatrix}      1 & 0 &0\\      0 & 1 & 0\\      0 & 0 &0      \end{bmatrix}  \begin{bmatrix} \hat{g}_x\\ \hat{g}_y\\ \hat{g}_z\end{bmatrix},
\end{eqnarray}
that projects the sensed vertical $\hat{\bm{g}}$ into the x-y plane in the head coordinate system, eliminating the value on the z-axis.
Finally, $\bm{\hat{g}}$ can be updated by
\begin{eqnarray}
\frac{d\hat{\bm{g}}}{dt}=\bm{K}_{vvc}~(\bm{vv}_{s}-\bm{\hat{vv}}_{s} )+\bm{K}_{vc}(\bm{v}_s-\bm{\hat{v}}_s). 
\end{eqnarray}

\subsection{Motion Sickness Estimation}

According to the SVC theory~\citep{bles1998motion}, motion sickness is mainly caused by a conflict between vertical perception through sensory organs and vertical feeling estimated by their internal models.
Therefore, \citet{bos1998modelling} proposed that motion sickness incidence~(MSI), which represents the percentage of vomiting subjects, is determined by the conflict of vertical signals $\mathit{\Delta} \bm{v}=\bm{v}_s-\hat{\bm{v}}_s$ using
\begin{eqnarray}
MSI = \frac{P}{(\tau_l s+1)^2} \frac{||\mathit{\Delta} \bm{v}||/b}{1+||\mathit{\Delta} \bm{v}||/b},
\end{eqnarray}
where $\frac{||\mathit{\Delta} \bm{v}||/b}{1+||\mathit{\Delta} \bm{v}||/b}$ is the Hill function that normalizes the L2 norm of the vertical conflict signal $||\mathit{\Delta} \bm{v}||$ to [0,1).

\section{DRIVING EXPERIMENT}

This experiment aimed to verify whether the proposed 6~DoF~SVC-VV model can predict MSI while riding an APMV with different visual conditions.
Therefore, two APMV ride comparison conditions were established for this experiment: \  1) looking ahead during autonomous driving (LAD), and 2) working with a tablet device during autonomous driving (WAD).
Considering that the tablet device used in WAD may hinder passengers' visual-spatial perception, we propose the following hypothesis:
\begin{itemize}
    \item[\textbf{H~1}:]  Passengers working while riding the APMV will have a higher probability of getting motion sickness than if they look ahead while riding the APMV.
\end{itemize}
 
To simulate the use of APMV in stations or shopping malls where frequent pedestrian avoidance is required, an experiment in which participants rode an APMV was conducted in an indoor room environment. 
This study was conducted with the approval of the Research Ethics Committee of the Nara Institute of Science and Technology (No. ~2021-I-38).

\subsection{Autonomous personal mobility vehicle} \label{sec:APMV}

 \begin{figure}[b] 
\centering 
\includegraphics[width=1\linewidth]{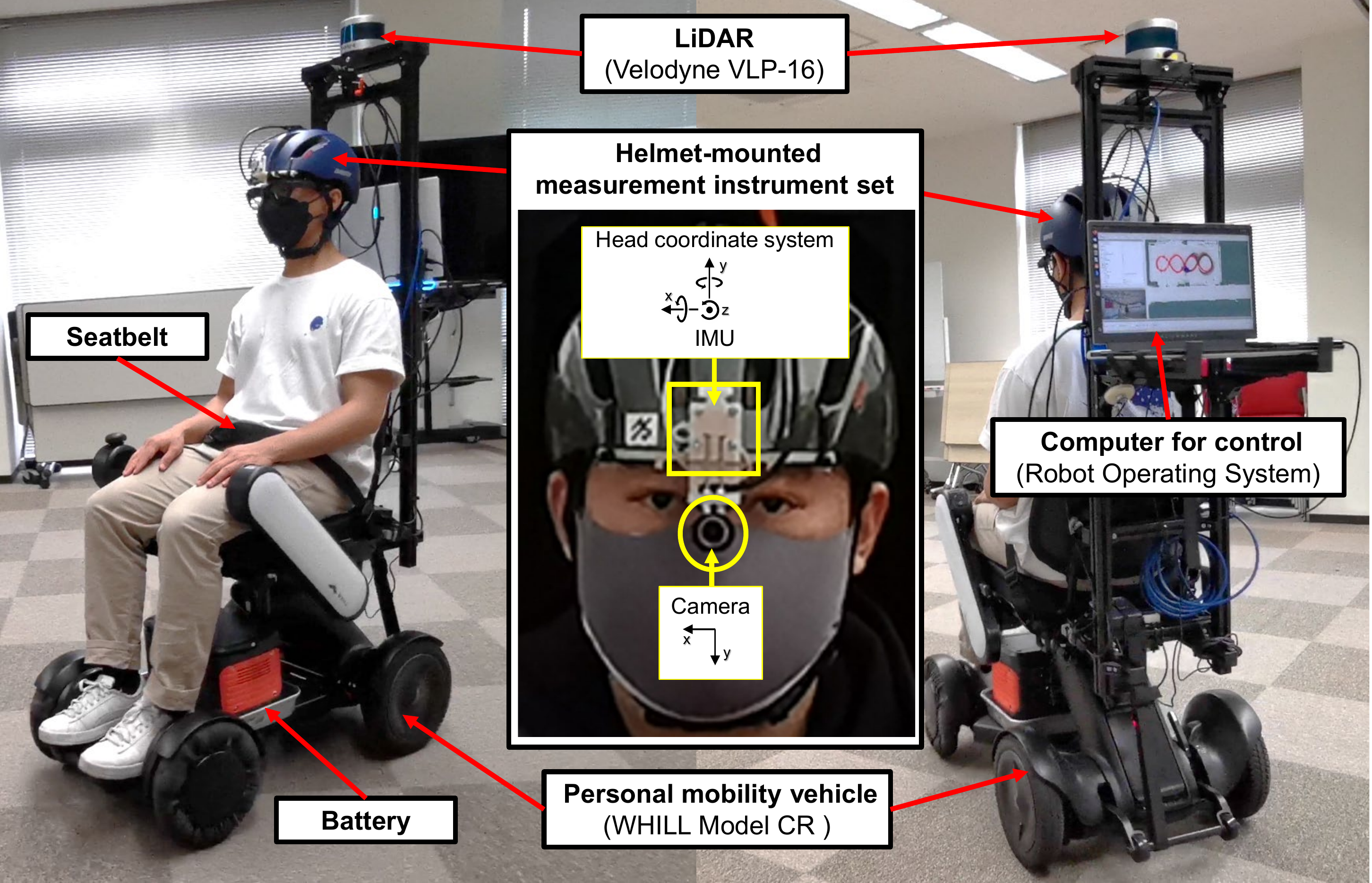}
\caption{Autonomous driving robotic wheelchair used as the experimental vehicle. A helmet-mounted measurement instrument~(HMMI): an IMU and a camera were placed on a helmet to observe the acceleration and angular velocity of the passenger's head and visual information.} 
\label{fig:WHILL}
\end{figure}
 
In this experiment, a robotic wheelchair {\it WHILL Model CR} with an autonomous driving system was used as the APMV.
As shown in Fig.~\ref{fig:WHILL}, the APMV was equipped with multilayered LiDAR (Velodyne VLP-16) and a controlling laptop PC.
An autonomous driving system based on the \textit{Robot Operating System} was applied to the APMV.
LiDAR was utilized for self-localization by the \textit{adaptive Monte Carlo localization} method on a previously built environmental map using the \textit{simultaneous localization and mapping} method.
Thus, it could automatically drive on pre-designed routes using a path-following controller~\citep{watanabe2016neonavigation}.

To ensure the experiment safety, the APMV had an automatic brake function that was applied when there was an obstacle within 0.5 meters directly a front of it.
Passengers could also actively control the APMV with the on-board joystick and power button if they feel in danger.
Meanwhile, a wireless remote controller could control the APMV to stop based on the actual risks during the experiment.
Further, the maximum velocity was set to $6~[km/h]$, and the maximum linear acceleration was set to $1.7~[m/s^2]$.

\subsection{Driving conditions}

\begin{figure}[!ht] 
\centering 
\includegraphics[width=0.75\linewidth]{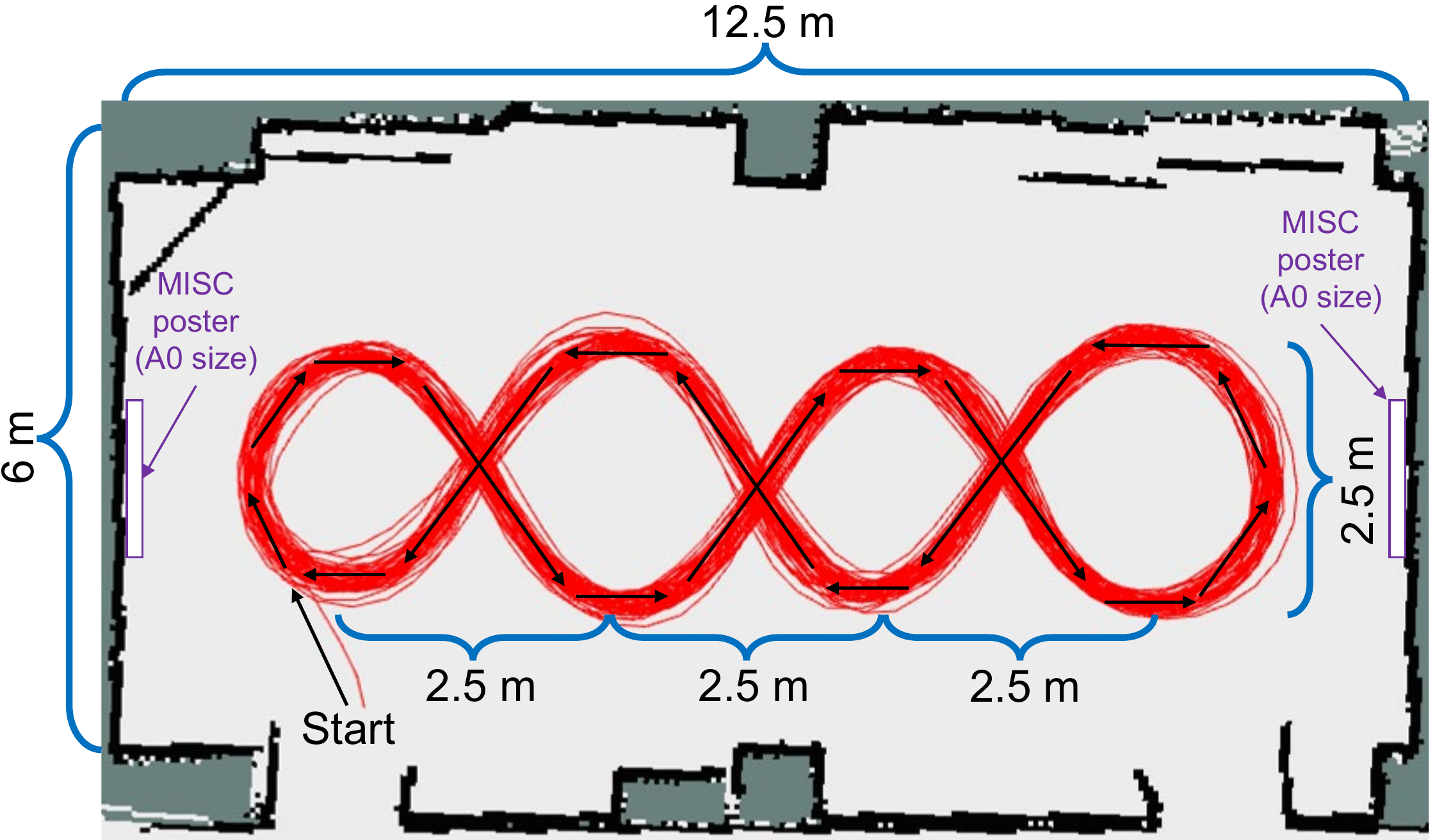}
\caption{$12.5~[m] \times 6~[m]$ room used as the experimental environment. A slalom path was used for autonomous driving. Two posters showing 11 MISC points (see the middle of Fig.~\ref{fig:exp01_MISC}) were mounted on the walls at either end of the room.}
\label{fig:Driving_path}
\end{figure}

As shown in Fig.~\ref{fig:Driving_path}, a $6~[m] \times 12.5~[m]$ room at Nara Institute of Science and Technology was used as the experimental site.
The target temperature of the air conditioner in the room was set as \ang{25}C.
In this experiment, a 20-min slalom driving path (see the red line in Fig.~\ref{fig:Driving_path}) was designed to simulate APMV avoiding other traffic participants in mixed traffic, such as shared space.
Specifically, APMV performed slalom driving with four centers of rotation such that the number of left and right rotations was the same.
The diameter of each rotation was approximately $2.5~[m]$; therefore, the distance between the two rotary centers was also $2.5~[m]$.
To reduce the effect of the participants' predictions of driving dynamics on their motion sickness, no actual object was placed in these centers of rotation.
 
\subsection{Riding conditions}
 
\begin{figure}[t] 
\centering 
\includegraphics[width=1\linewidth]{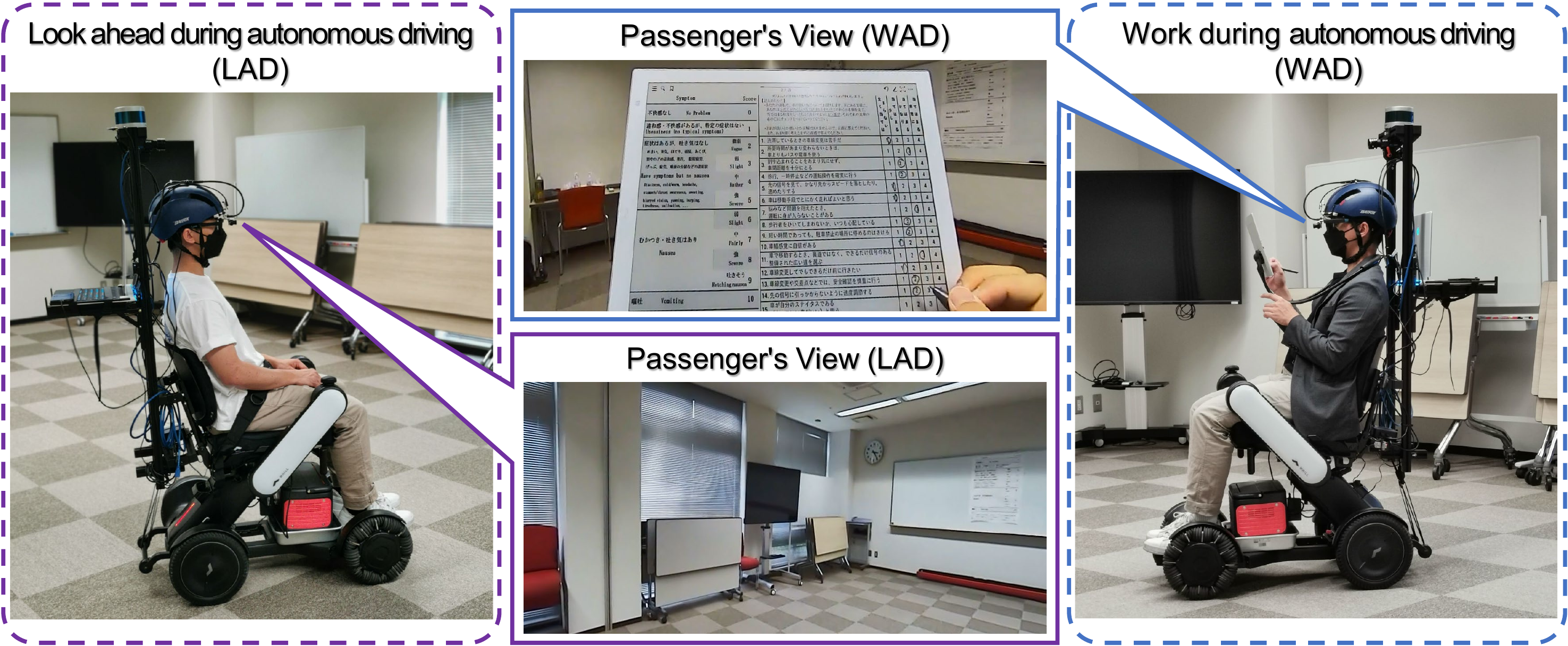}
\caption{Two riding conditions: 1) looking ahead during autonomous driving (LAD); 2) working with a tablet device during autonomous driving (WAD).} 
\label{fig:exp02_conditions}
\end{figure}

As shown in Fig.~\ref{fig:exp02_conditions}, two riding conditions were designed: 1) looking ahead during autonomous driving (LAD) and 2) working with a tablet device during autonomous driving (WAD).
Each scenario took 25 min, including 20 minuts of riding the APMV and 5 minuts of resting on the stopped APMV.
The detailed design of each scenario is as follows.

\subsubsection{Looking ahead during autonomous driving (LAD)}

In the LAD, participants were asked to look ahead during autonomous driving.
It was also hypothesized that participants could easily obtain vertical orientation information from the floor, walls, and surrounding objects such as windows, tables, chairs, and whiteboards.
 
\subsubsection{Working with a tablet device during autonomous driving (WAD)}

In the WAD, participants were asked to work with a tablet device (Sony DPT-RP1 Digital Paper:224 [mm] high with 302.6 [mm] weight for horizontal use).
They were asked to use a stylus to answer dummy questionnaires and read articles on e-books while riding APMV.
Note that the contents of these questionnaires and articles were not relevant to this experiment to avoid influencing the experimental results.
Moreover, a neck-hanging tablet stand was used to help participants hold the e-book, see the text clearly, and write more easily.
It also enabled the relative positions of the head and tablet device to be maintained within a certain range.

This scenario was also hypothesized to be more prone to cause motion sickness because the participants may experience difficulty in recognizing the vertical direction because the tablet device prevented the passenger from perceiving the body motion from dynamic visual information~\citep{Sato2022} such as optical flow, and static visual information such as horizontal or vertical.

\subsection{Measurements}
\subsubsection{Head movement and visual information}

To measure the acceleration and angular velocity of the passenger's head and visual information, a helmet-mounted measurement instrument(HMMI) was used (see the center of  Fig.~\ref{fig:WHILL}).
The HMMI included an inertial measurement unit(IMU) and a camera set in front of a helmet.
The IMU measured 3 DoF acceleration $f$ and 3 DoF angular velocity $\omega$ at 100~[Hz].
The camera resolution was set to $1280\times 720$ pixels at 30~[Hz].
To reduce the impact of lens distortion on the visual vertical prediction, we cropped the periphery of the camera video.
The cropped resolution was $1000\times 480$.

\subsubsection{Motion sickness}

\begin{figure}[ht] 
\centering 
\includegraphics[width=0.45\linewidth]{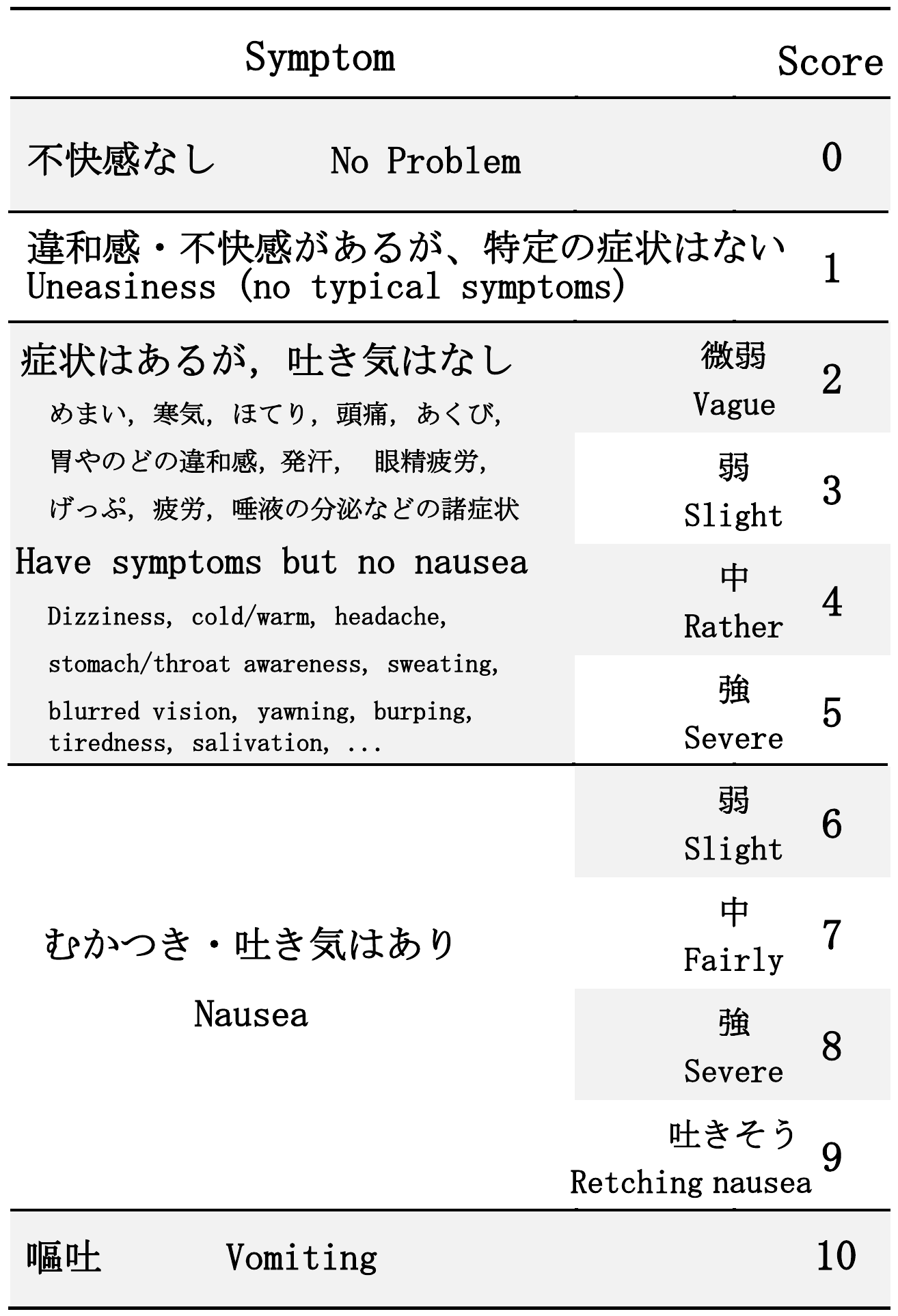}
\caption{An 11-point MIsery SCale (MISC)~\citep{MISC} in Japanese and English.} 
\label{fig:exp01_MISC}
\end{figure}

To measure the severity of motion sickness during the 25-minute experiment, participants verbally reported their feelings of motion sickness every minute using an 11-point MIsery~SCale (MISC)~\citep{MISC} ranging from zero to 10 (see Fig.~\ref{fig:exp01_MISC}).
If MISC reached 6 and lasted for more than 2 minutes, then the APMV stopped, and participants continued to sit on the stopped APMV to report MISC every minute for 5 min. 
To help participants refer to the definition of MISC, the two A0 size posters in Fig.~\ref{fig:exp01_MISC} were placed on whiteboards on both sides of the room in the LAD scenario; and the definition of MISC was available on each page of the e-book in the WAD scenario (see Fig.~\ref{fig:exp02_conditions}).

\subsection{Participants and groups}
 
A total of 27 participants (10 females and 17 males) participated in this experiment as users of APMV.
They were 22-29 years old (mean:23.5, standard deviation,1.89). 
They had no experience with autonomous cars and APMVs before this experiment.
All participants provided informed consent before participating in the experiment.
Each participant was asked to ride the APMV under the LAD and WAD conditions once.

To avoid the order effect of the experimental conditions on the experimental results, participants were randomly assigned to two groups.
Specifically, 14 participants in a group called \textit{LAD$\rightarrow$WAD} experienced the LAD scenario first and then the WAD scenario.
Thirteen participants in another group called \textit{WAD$\rightarrow$LAD} experienced these conditions in the opposite order.

The minimum time interval between the two conditions was 24 hours.
The total time of this experiment, including the two conditions was four hours for each participant. 
Each participant received 4,000 Japanese Yen as a reward.

\subsection{Procedure}
 
First, the participants were instructed by following information before the experiment:
\begin{itemize}
       \item The purpose of the experiment was to investigate the effect of the riding conditions of an APMV on motion sickness.
    \item In this experiment, participants rode on the APMV and experienced slalom driving repeatedly (up to 6 km/h).
    \item There were two riding conditions. Each riding scenario was performed on a separate day. Each scenario took approximately 2 hours, and the total time for the two conditions was approximately 4 hours.
    \item Details of those two riding conditions were introduced just before participants experienced one of them, separately.
    \item In each riding scenario, the APMV was autonomously driven in 20 minutes, then the APMV stopped, and participants could rest for 5 minutes on it.
    \item In those 25 minutes, participants were required to report a level of MISC in each minute based on their feelings of motion sickness.
    \item If the MISC reached 6 and lasted for more than 2 minutes, then the APMV was stopped immediately and participants continued to sit on the stopped APMV to report MISC every minute for 5 minutes. 
   \end{itemize}
 
Moreover, to reduce the restlessness and nervousness of the participants owing to lack of knowledge about APMV, we explained the principles of the autonomous driving system and its sensors (\ie Lider), and the operational design domain (\eg sensor range, maximum speed, maximum linear acceleration, and the judgment distance of emergency stop) to the participants in detail.
Participants were allowed to actively take over and stop the APMV if they thought there was danger.

\subsection{Evaluation methods}

\subsubsection{Motion sickness symptoms reported by MISC}
 
The MISC was reported by each participant every minute during each 25-min trial.
The mean and maximum of the MISC in each trial were counted to evaluate the degree of motion sickness of the participants.
 
We used $2\times2$ mixed-design ANOVAs to evaluate the mean and maximum of MISC in two riding conditions (within-subject factor: LAD and WAD) between two groups of condition order (between-subject factor:  \textit{LAD$\rightarrow$WAD} and  \textit{WAD$\rightarrow$LAD}).
Furthermore, MISC results were also analyzed under LAD and WAD conditions to test whether our proposed hypothesis \textbf{H~1}, \ie passengers working with a tablet device while riding the APMV will have a higher probability of motion sickness than if they look ahead while riding the APMV.

\subsubsection{Calculated visual vertical}
 
The calculated visual vertical of each trial was evaluated by analyzing the Pearson correlation coefficient between the direction of VV, \ie $\theta^{vv}$, and direction of gravitational acceleration, \ie $\theta^{g}$, from each 25-min trial independently.
Particularly, $\theta^{vv}$ was estimated using Algorithm~\ref{VV} from the camera data, and the direction of the gravitational acceleration projected in the 2D head coordinate system was calculated as $\theta^{g}=180~\arctan(g_y/g_x)/\pi$.
Here, the gravitational acceleration $\bm{g}$ was estimated from the IMU data, that is, $\bm{f}$ and $\bm{\omega}$.
Moreover, a two-sided paired t-test was used to analyze the significant difference between the Pearson correlation coefficients for the LAD and WAD conditions.


\subsubsection{Motion sickness prediction by MSI}

The proposed 6~DoF~SVC-VV model was used to predict the MSI from the IMU data (\ie $\bm{f}$ and $\bm{\omega}$) and the camera images measured in the experiment. To implement the calculation of the zero-order holder, the calculated visual vertical was up-sampled from 30 [Hz] to 100 [Hz] to synchronize with the IMU data.

A conventional 6~DoF SVC model proposed by \citep{Inoue2022} (called the In1 model in this study) and its optimized parameters were used as a baseline.
Furthermore, the parameters of the 6~DoF~SVC-VV model were the same as those of the 6~DoF SVC model, \ie In1 model in~\citep{Inoue2022}, except for $K_{vc}$ and a new parameter $K_{vvc}$.
Table~\ref{tab:parameter} lists the parameters used in the two models.
Note that the conventional 6~Dof SVC model (In1) can be described using the proposed 6 DoF-SVC-VV model if the parameters are set as $K_{vc}=5.0$ and $K_{vvc}=0.0$. Therefore, 
for the 6~DoF~SVC-VV model, parameters $K_{vvc}=2.5$  $K_{vc}=2.5$ were used to balance the feedback strength of the two conflict signals.
 
Similar to the MISC evaluation method, $2\times2$ mixed-design ANOVAs (within-subject factors: LAD and WAD; between-subject factors:  \textit{LAD$\rightarrow$WAD} and  \textit{WAD$\rightarrow$LAD}) were used to evaluate the mean and maximum values of the predicted MSI in LAD and WAD, respectively. 

 \begin{figure}[!ht]
\centering
\captionof{table}{Parameters for the 6 DoF SVC model and the 6 DoF SVC-VV model.}
\label{tab:parameter}
\setstretch{1.3}
\begin{tabular}{@{}ccccccccccccc@{}}
\toprule
Model & $K_a$ & $K_{\omega}$ & $K_{\omega c}$ & $K_{ac}$ & $K_{vc}$ & $K_{vvc}$ & \begin{tabular}[c]{@{}c@{}}$\tau$\\ $[s]$\end{tabular} & \begin{tabular}[c]{@{}c@{}}$\tau_{a}$\\$[s]$\end{tabular}  & \begin{tabular}[c]{@{}c@{}}$\tau_{d}$\\ $[s]$\end{tabular} & \begin{tabular}[c]{@{}c@{}}$b$\\ $[m/s^2]$\end{tabular} & \begin{tabular}[c]{@{}c@{}}$\tau_{I}$\\ $[s]$\end{tabular} & \begin{tabular}[c]{@{}c@{}}$P$\\ $[\%]$\end{tabular} \\ \midrule
6~DoF SVC & 0.1 & 0.1 & 10 & 0.5 &\textbf{5.0} & \textbf{0.0} & 2.0 &190.0 & 7.0 & 0.5 &  720.0 & 85 \\ 
6~DoF~SVC-VV & 0.1 & 0.1 & 10 & 0.5 &\textbf{2.5} & \textbf{2.5} & 2.0 & 190.0 & 7.0 & 0.5 & 720.0 & 85 \\ \bottomrule
\end{tabular}
\end{figure}

\subsubsection{Comparison between predicted MSI and reported MISC}
 
To investigate the performance of the proposed 6 DoF SVC-VV model, we compared the predicted MSI with the reported MISC.
Note that MSI and MISC are different indicators for each other; \ie MSI indicates the percentage of participants who experienced vomiting when exposed to motion for a certain time, and MISC indicates the subjective assessment of each participant of the severity of motion sickness.
Basically, the MISC is an evaluation indicator for individuals, whereas the MSI is an evaluation indicator for the whole group.
Considering the difference in meaning between MSI and MISC, the performance of our proposed 6 DoF SVC-VV model was evaluated by comparing the high-low relationship of the reported MISC from each participant under the LAD and WAD, and that of the predicted MSI under those two riding conditions.
 
Based on the confusion matrix presented in Table~\ref{tab:TP_FP}, multiple evaluation indexes, that is, accuracy, precision, recall, and F1 score (Table~\ref{tab:evaluation_index}), were used to evaluate the performance of the proposed 6 DoF SVC-VV model.
We took each participant's reported MISC as the true result and the predicted MSI as the predicted result of their motion sickness.
Therefore, as summarized in Table~\ref{tab:TP_FP}, we refer to \textbf{H~1}, ${MISC}_{LAD} < {MISC}_{WAD}$ and  ${MISC}_{LAD} \geq {MISC}_{WAD}$ as the positive and negative states of the true result, respectively.
Meanwhile, ${MSI}_{LAD} < {MSI}_{WAD}$ and ${MSI}_{LAD} \geq {MSI}_{WAD}$ are considered the positive and negative states of the predicted result, respectively.
Moreover, the mean and maximum values were used as representative values for MISC and MSI.

\begin{figure}[t]
\centering
\captionof{table}{Definition of confusion matrix for evaluating motion sickness prediction by the predicted MSI comparing to the reported MISC.}
\label{tab:TP_FP}
\setstretch{1.2}
\begin{tabular}{@{}cccc@{}}
\toprule
&  & \multicolumn{2}{|c}{\begin{tabular}[c]{@{}c@{}}MISC\\ (True result)\end{tabular}} \\ \cmidrule(l){3-4} 
& \multicolumn{1}{c|}{} & \multicolumn{1}{c|}{\begin{tabular}[c]{@{}c@{}}Positive\\ ($MISC_{LAD} < MISC_{WAD}$)\end{tabular}} & \begin{tabular}[c]{@{}c@{}}Negative\\ ($MISC_{LAD} \geq MISC_{WAD}$)\end{tabular} \\ \midrule
\multirow{2}{*}{\begin{tabular}[c]{@{}c@{}}\\MSI\\ (Predicted result)\end{tabular}} & \multicolumn{1}{|c|}{\begin{tabular}[c]{@{}c@{}}Positive\\ ($MSI_{LAD} < MSI_{WAD}$)\end{tabular}} & \multicolumn{1}{c|}{\begin{tabular}[c]{@{}c@{}}TP\\ (True Positive)\end{tabular}} & \begin{tabular}[c]{@{}c@{}}FP\\ (False Positive)\end{tabular} \\ \cmidrule(l){2-4} 
& \multicolumn{1}{|c|}{\begin{tabular}[c]{@{}c@{}}Negative\\ ($MSI_{LAD}\geq MSI_{WAD}$)\end{tabular}} & \multicolumn{1}{c|}{\begin{tabular}[c]{@{}c@{}}FN\\ (False Negative)\end{tabular}} & \begin{tabular}[c]{@{}c@{}}TN\\ (True Nagative)\end{tabular} \\ \bottomrule
\end{tabular}
\end{figure}

\begin{figure}[t]
\centering
\centering
\captionof{table}{Evaluation indexes for the prediction of motion sickness using predicted MSI compared to reported MISC.}
\label{tab:evaluation_index}
\setstretch{2}
\begin{tabular}{@{}ll@{}}
\toprule
Evaluation index & Explanation \\ \midrule
$\displaystyle Accuracy = \frac{TP+TN}{TP+TN+FP+FN}$& Rate of the correct predictions over all predictions. \\
$\displaystyle Precision= \frac{TP}{TP+FP}$& Rate of correct positive predictions over all positive predictions. \\ 
$\displaystyle Recall=\frac{TP}{TP+FN}$& Rate of correct positive predictions over all the positive true results. \\ 
$\displaystyle F1~score=\frac{2\cdot Precision \cdot Recall}{Precision+Recall}$ & The harmonic mean of the precision and recall. \vspace{2mm}\\
\bottomrule
\end{tabular}
\end{figure}

\section{RESULTS}

\subsection{Reported MISC}

\begin{figure}[t] 
\centering 
     \begin{subfigure}[b]{0.495\textwidth}
         \centering
         \includegraphics[width=0.7\textwidth]{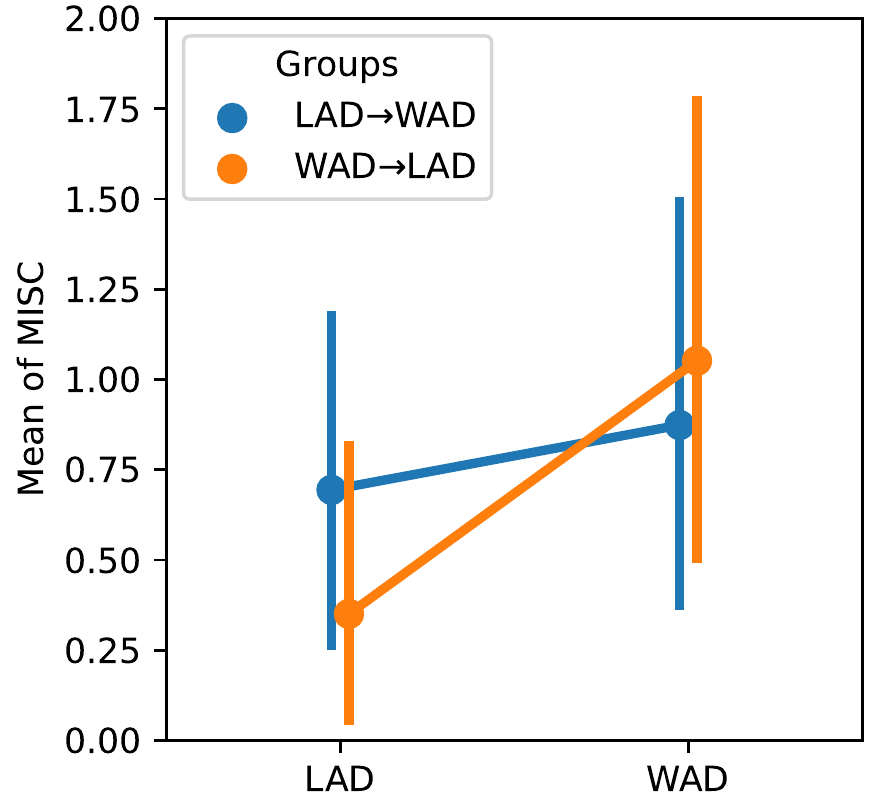}
         \caption{Mean of MISC}
         \label{fig:MISC_mean_max_a}
     \end{subfigure}
     \hfill
     \begin{subfigure}[b]{0.495\textwidth}
         \centering
         \includegraphics[width=0.7\textwidth]{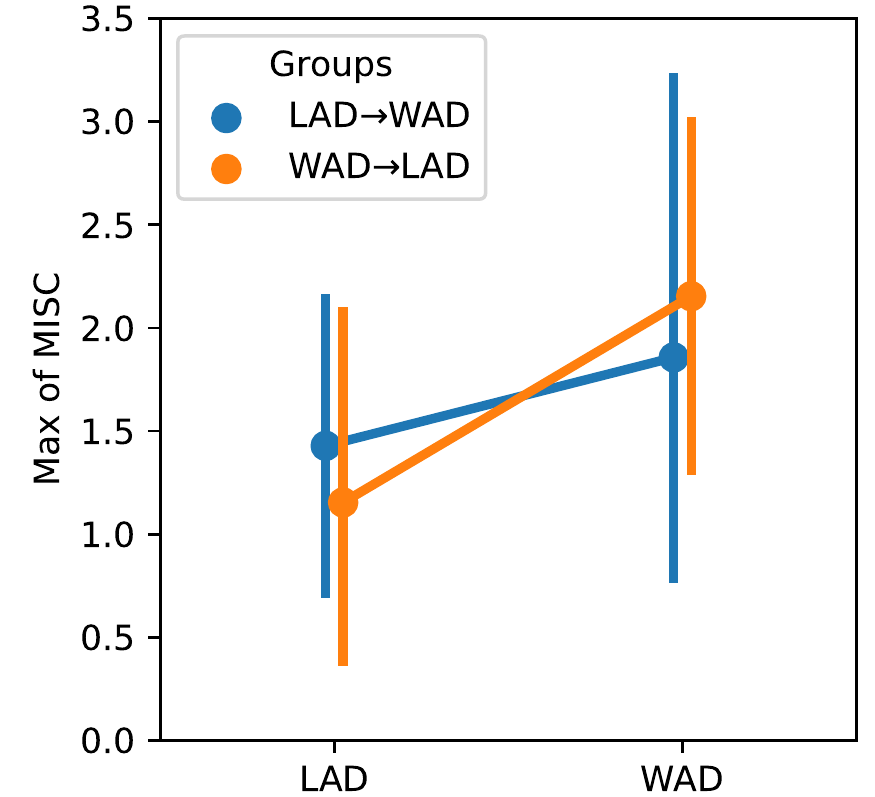}
         \caption{Maximum of MISC}
         \label{fig:MISC_mean_max_b}
     \end{subfigure}
\captionof{figure}{Mean and Maximum of MISC reported under LAD and WAD conditions (error bar: $95\%$ confidence interval)} 
\label{fig:MISC_mean_max}
\centering
\captionof{table}{Two-way mixed-design ANOVA for mean and maximum of MISC. * shows the $p<.05$}
\label{tab:MISC_ANOVA}
\setstretch{1.3}
\begin{tabular}{rrrrrrrlrr}
\toprule
\multicolumn{1}{c}{Measurement} &\multicolumn{1}{c}{Effect} & \multicolumn{1}{c}{SS} & \multicolumn{1}{c}{DF1} & \multicolumn{1}{c}{DF2} & \multicolumn{1}{c}{MS} & \multicolumn{1}{c}{\textit{F-value}} & \multicolumn{1}{c}{\textit{p-value}} & \multicolumn{1}{c}{np2} & \multicolumn{1}{c}{eps} \\  \midrule
Mean of MISC & Groups & 0.092 & 1 & 25 & 0.092 & 0.065 & 0.801 & 0.003 & NaN \\
& Conditions & 2.506 & 1 & 25 & 2.506 & 4.296 & \textbf{0.049 *} & 0.147 & 1.0 \\
& Interaction & 0.918 & 1 & 25 & 0.918 & 1.574 & 0.221 & 0.059 & NaN \\  \midrule
Maximum of MISC & Groups & 0.002 & 1 & 25 & 0.002 & 0.000 & 0.986 & 0.000 & NaN \\
& Conditions & 6.685 & 1 & 25 & 6.685 & 3.362 & 0.079 & 0.119 & 1.0 \\
& Interaction & 1.101 & 1 & 25 & 1.101 & 0.553 & 0.464 & 0.022 & NaN \\  \bottomrule
\end{tabular}
\end{figure}

\begin{figure}[!hp]
\vspace{3mm}
\centering 
\includegraphics[width=1\linewidth]{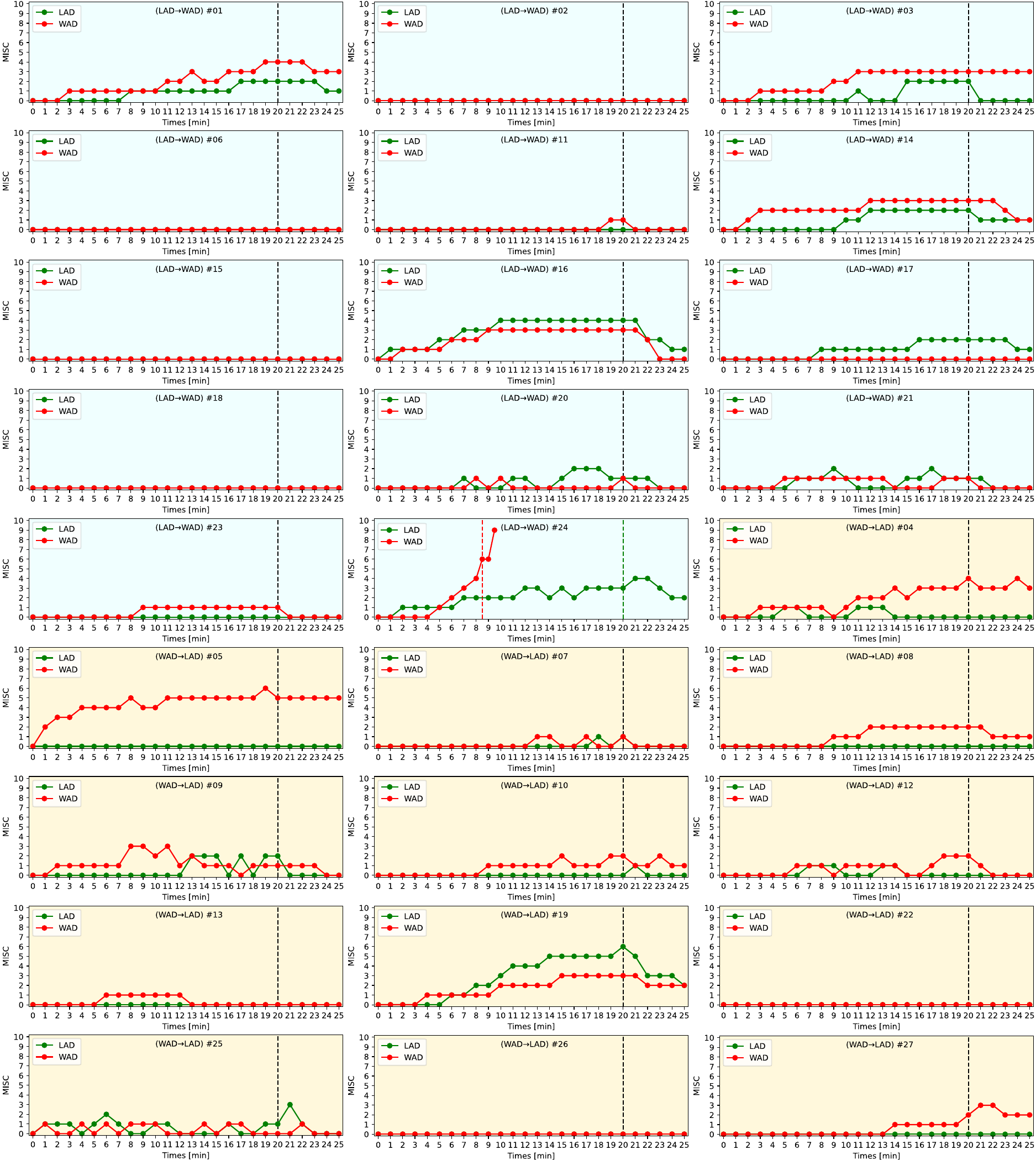}
\caption{MISC reported from 27 participants every minute. The horizontal coordinate shows the time whereas vertical coordinate shows the MISC. Graphs on the sky blue background represent MISCs reported by participants from \textit{LAD$\rightarrow$WAD} group whereas graphs on the yellow background represent MISCs reported by participants from \textit{WAD$\rightarrow$LAD} group. Green lines represent MISC reported by participants in LAD whereas red lines represent MISC reported by participants in WAD. The vertical broken lines indicate the moment of APMV stopping.}  
\label{fig:MISC_TS_all}
\end{figure}

The MISC per minute reported for 27 participants is shown in Fig.~\ref{fig:MISC_TS_all}. 
Among them, 21 participants reported that they developed symptoms of motion sickness in this experiment; however, six participants (four in group \textit{LAD$\rightarrow$WAD} and two in group \textit{WAD$\rightarrow$LAD}) reported $MISC=0$ at all times in both the LAD and WAD conditions. 
In the WAD, participant \#24 reported that $MISC=6$ at approximately 8.5 minutes and was asked to stop the APMV. 
Then, the APMV was stopped, and the participant rested on the APMV. 
However, after 1 minut of rest, the motion sickness symptoms of participant \#24 continued to develop to $MISC=9$ at approximately 9.5 minutes, thus we immediately terminated the experiment.

For the two groups, \ie, \textit{LAD$\rightarrow$WAD} and \textit{WAD$\rightarrow$LAD}, we did not find a significant effect of the experimental order on the MISC results, as summarized in Table~\ref{tab:MISC_ANOVA}, and the mean and maximum of MISC during 25 minuts driving in LAD were lower than those in WAD, as shown in Fig.~\ref{fig:MISC_mean_max}. 

For two conditions, \ie LAD and WAD, the two-way mixed-design ANOVA (see Table~\ref{tab:MISC_ANOVA}) reported that there was a significant difference in the mean MISC between conditions ($p=0.049$); however, no significant difference between groups and in their interaction. 
Moreover, there was no significant difference in the maximum MISC between the groups, conditions, and their interactions.

\subsection{Calculated visual vertical}

\begin{figure}[!hb] 
\centering  
\includegraphics[width=0.65\linewidth]{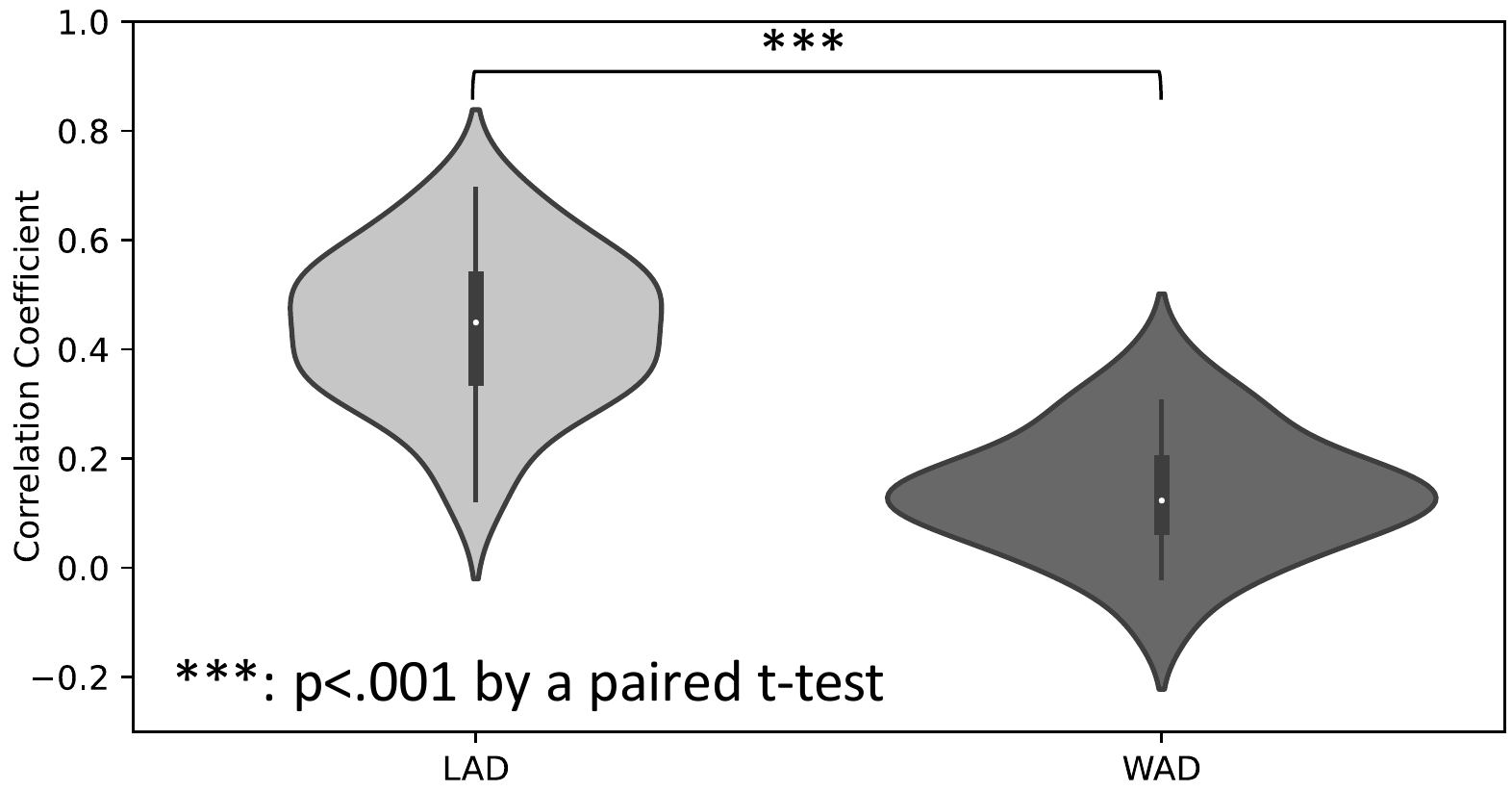} 
\caption{Pearson correlation between the directions of the predicted visual vertical and estimated gravitational acceleration. A two-sided paired t-test reports a significant difference between the Pearson correlation coefficients under LAD and WAD conditions.}  \label{fig:Corr_LAD_WAD} 
\end{figure}

Fig.~\ref{fig:Corr_LAD_WAD} shows that $\theta^{vv}$ and $\theta^{g}$ had positive Pearson correlations ($N=27, M = 0.44, SD = 0.14$) under the LAD condition, \ie without the obstruction of view, and weakly positive Pearson correlations under the WAD condition, \ie with obstruction of view.
Furthermore, a two-sided paired t-test showed that the mean value of the Pearson correlation coefficients in the LAD condition was significantly higher than that in the WAD condition, \ie, $t(26) = 11.11, p < .001$. 

\begin{figure}[!hp] 
\vspace{3mm}
\centering 
\includegraphics[width=1\linewidth]{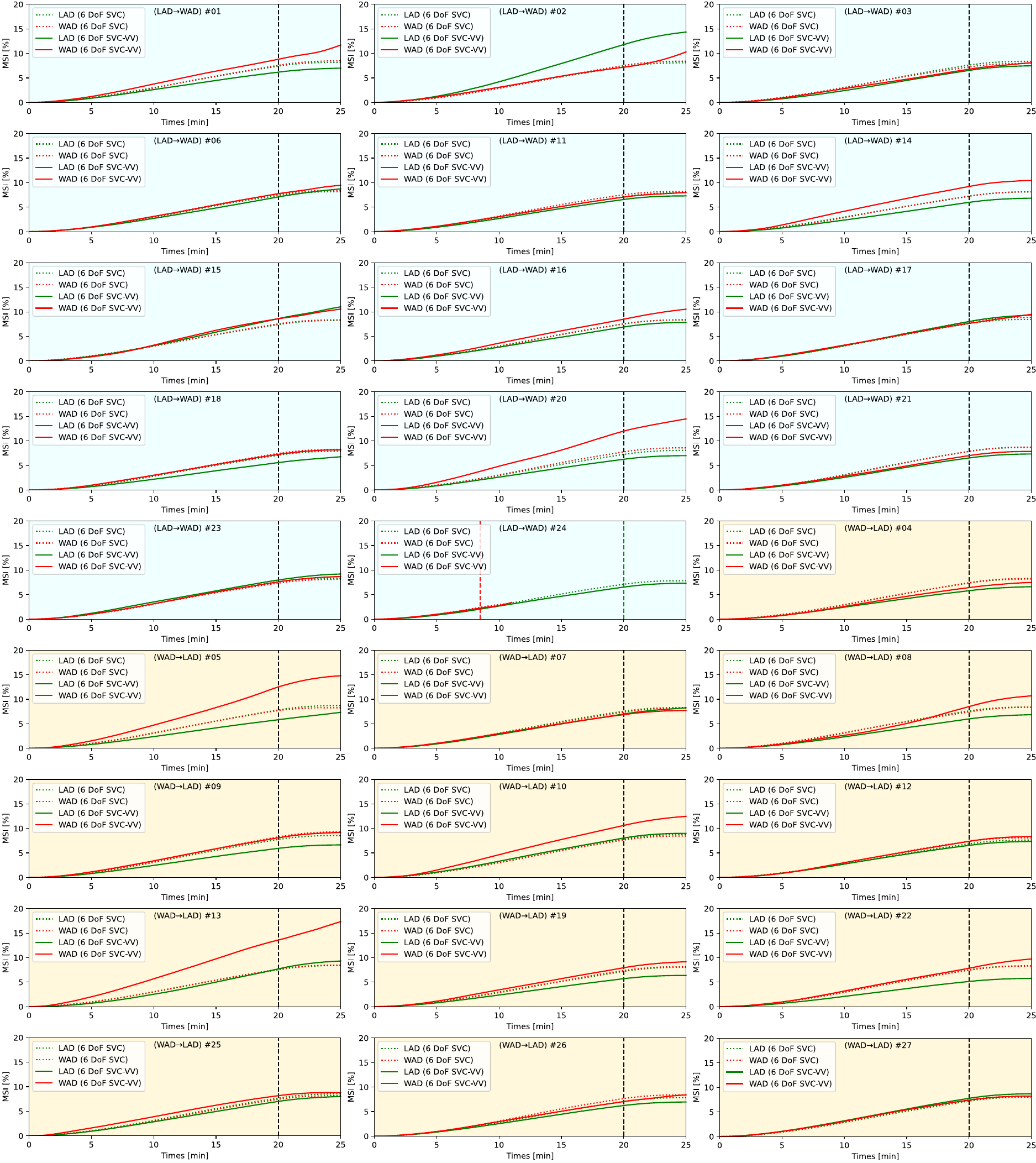}
\caption{Predicted MSI during the 25 minuts experiment by 6 DoF SVC and 6 DoF SVC-VV models for 27 participants. The horizontal coordinate is the time and the vertical coordinate is the vomiting rate, \ie MSI. Graphs on the sky blue background represent MSIs predicted for participants from \textit{LAD$\rightarrow$WAD} group whereas graphs on the yellow background represent MSIs predicted for participants from \textit{WAD$\rightarrow$LAD} group.  Green lines represent MISC reported by participants in LAD whereas red lines represent MISC reported by participants in WAD. Dotted lines represent MSI predicted by 6 DoF SVC model whereas solid lines represent MSI predicted by 6 DoF SVC-VV model. The vertical broken lines indicate the APMV stopping moment.} 
\label{fig:MSI_TS_all}
\end{figure}

\newpage
\subsection{Predicted MSI}

\begin{figure}[b] 

     \centering
     \begin{subfigure}[b]{0.245\textwidth}
         \centering
         \includegraphics[width=\textwidth]{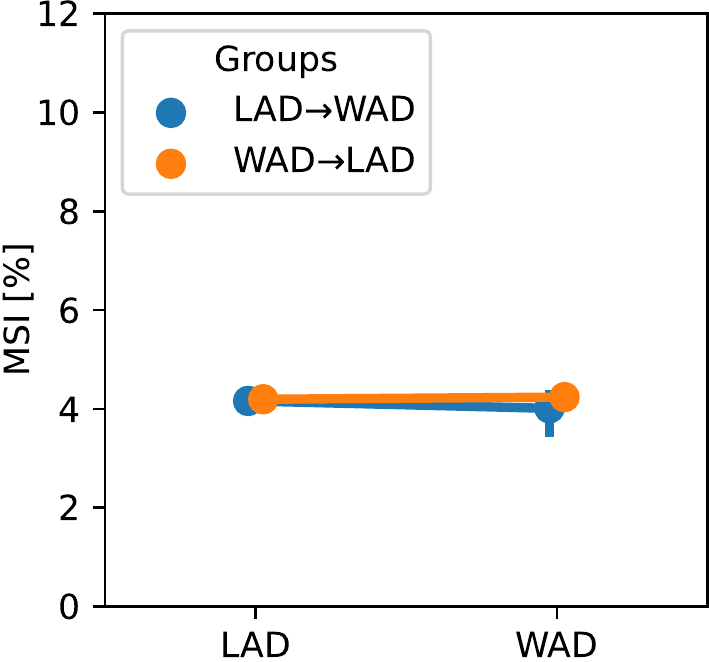}
         \caption{Mean of MSI by\\6 DoF SVC model}
         \label{fig:MSI_mean_max_a}
     \end{subfigure}
     \hfill
     \begin{subfigure}[b]{0.245\textwidth}
         \centering
         \includegraphics[width=\textwidth]{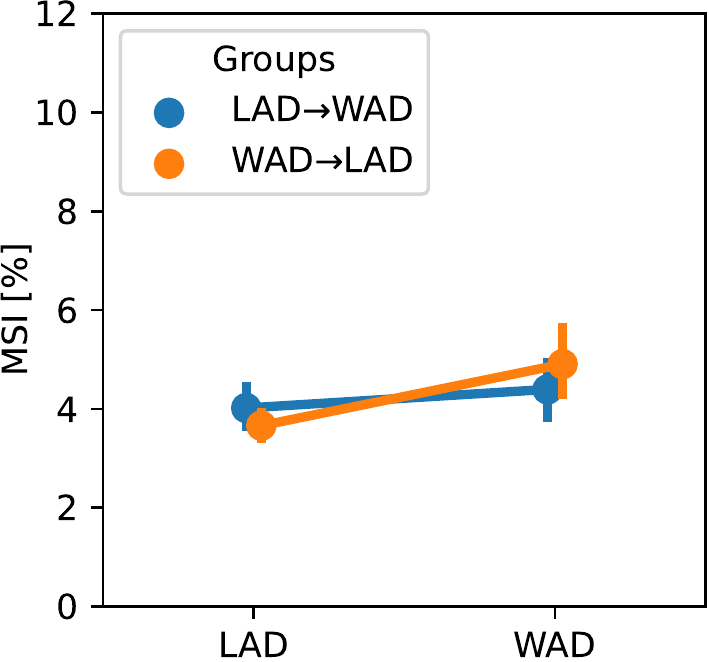}
         \caption{Mean of MSI by\\6 DoF SVC-VV model}
         \label{fig:MSI_mean_max_b}
     \end{subfigure}
     \hfill
     \begin{subfigure}[b]{0.245\textwidth}
         \centering
         \includegraphics[width=\textwidth]{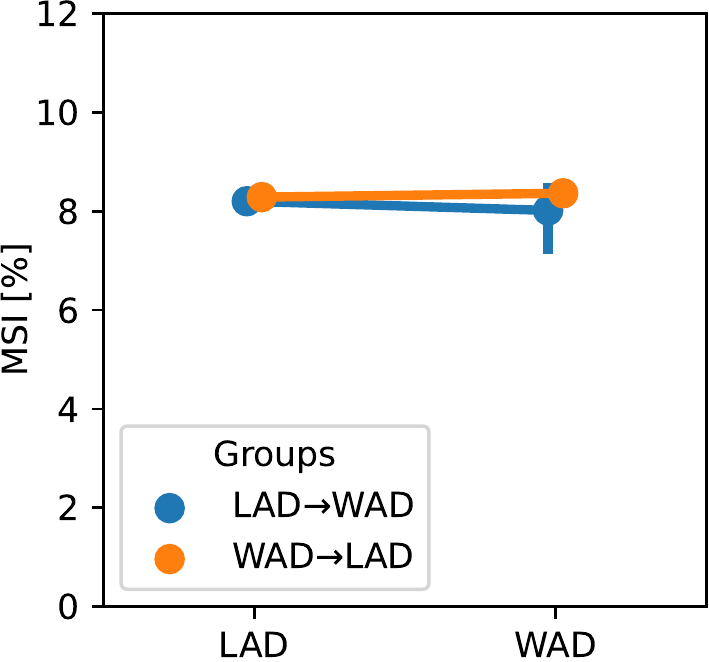}
         \caption{Maximum of MSI by\\6 DoF SVC model}
         \label{fig:MSI_mean_max_c}
     \end{subfigure}
       \hfill
     \begin{subfigure}[b]{0.245\linewidth}
         \centering
         \includegraphics[width=\linewidth]{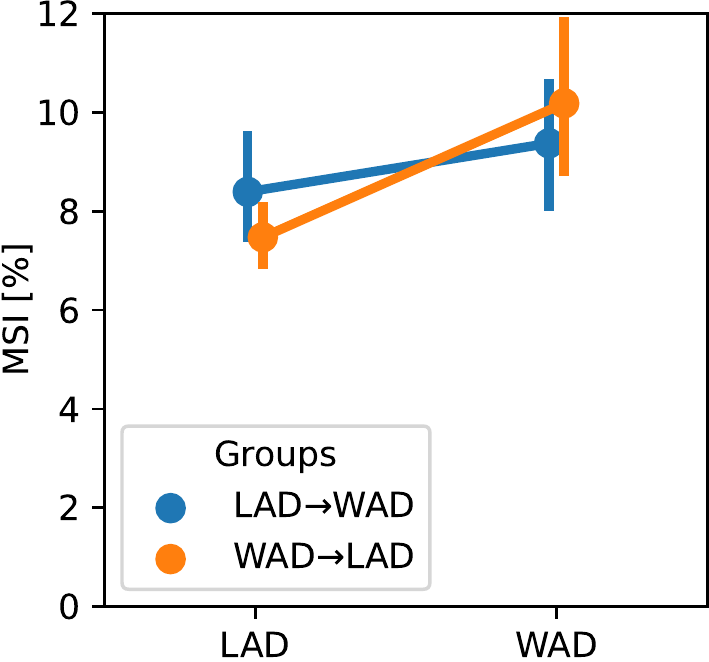}
         \caption{Maximum of MSI by\\6 DoF SVC-VV model}
         \label{fig:MSI_mean_max_d}
     \end{subfigure}
        \caption{Mean and maximum MSI predicted by the 6 DoF SVC model and the 6 DoF SVC-VV model (error bar: $95\%$ confidence interval).}
        \label{fig:MSI_resuts}
\centering
\captionof{table}{Two-way mixed-design ANOVA for mean of MSI predicted by 6 DoF SVC model and 6 DoF SVC-VV model, separately. ** shows the $p<.01$.}
\label{tab:ANOVA_Mean_MSI}
\begin{tabular}{rrrrrrrlrr}
\toprule
\multicolumn{1}{c}{Model} & \multicolumn{1}{c}{Source} & \multicolumn{1}{c}{SS} & \multicolumn{1}{c}{DF1} & \multicolumn{1}{c}{DF2} & \multicolumn{1}{c}{MS} & \multicolumn{1}{c}{\textit{F-value}} & \multicolumn{1}{c}{\textit{p-value}} & \multicolumn{1}{c}{np2} & \multicolumn{1}{c}{eps} \\ \midrule
6 DoF SVC model & Groups & 0.231  &  1  & 25 & 0.231 & 0.989 & 0.329 & 0.038 & NaN \\
& Conditions &  0.043  &  1 &  25 & 0.043 & 0.244 & 0.626 & 0.010 & 1.0 \\
& Interaction & 0.126 &   1 &  25 & 0.126 & 0.709 & 0.408 & 0.028 & NaN \\ \midrule
6 DoF SVC-VV model & Groups & 0.081  &  1  & 25 & 0.081 & 0.077 & 0.784 & 0.003 & NaN \\
& Conditions & 8.527  &  1 &  25 & 8.527 & 9.487 & \textbf{0.005 **} &  0.275 & 1.0 \\
& Interaction & 2.545   & 1 &  25 & 2.545 & 2.831 & 0.105&  0.102 & NaN \\ \bottomrule
\end{tabular}
\vspace{3mm}
\centering
\captionof{table}{Two-way mixed-design ANOVA for Maximum of MSI predicted by 6 DoF SVC model and 6 DoF SVC-VV model, separately. ** shows the $p<.01$.}
\label{tab:ANOVA_Max_MSI}
\begin{tabular}{rrrrrrrlrr}
\toprule
\multicolumn{1}{c}{Model} & \multicolumn{1}{c}{Source} & \multicolumn{1}{c}{SS} & \multicolumn{1}{c}{DF1} & \multicolumn{1}{c}{DF2} & \multicolumn{1}{c}{MS} & \multicolumn{1}{c}{\textit{F-value}} & \multicolumn{1}{c}{\textit{p-value}} & \multicolumn{1}{c}{np2} & \multicolumn{1}{c}{eps} \\ \midrule
6 DoF SVC model & Groups & 0.625  &  1 &  25 & 0.625 & 0.943 & 0.341 & 0.036 &  NaN \\
& Conditions & 0.048 &   1  & 25 & 0.048 & 0.104 & 0.750 & 0.004 & 1.0 \\
& Interaction & 0.234  &  1  & 25 & 0.234 & 0.511 & 0.481 &  0.020 & NaN \\ \midrule
6 DoF SVC-VV model & Groups & 0.042  &  1  & 25 &  0.042 &  0.007 & 0.935 & 0.000 & NaN\\
& Conditions & 44.45 &   1 &  25 & 44.45 & 10.72 & \textbf{0.003 **} & 0.300 & 1.0 \\
& Interaction & 10.07  &  1 &  25 & 10.07  & 2.429 & 0.132 & 0.089 & NaN \\ \bottomrule
\end{tabular}
\vspace{-3mm}
\end{figure}

The time series MSI predicted by the 6 DoF SVC and 6 DoF SVC-VV models are shown in Fig.~\ref{fig:MSI_TS_all}.

As a summarized index of the time-series MSI in Fig.~\ref{fig:MSI_TS_all}, the mean MSI predicted by the 6 DoF SVC and 6 DoF SVC-VV models are shown in Figs. ~\ref{fig:MSI_resuts}–\subref{fig:MSI_mean_max_a} and~\subref{fig:MSI_mean_max_b}, respectively.

As presented in Table~\ref{tab:ANOVA_Mean_MSI}, a two-way mixed-design ANOVA for the mean MSI predicted by the 6 DoF SVC model revealed no significant effect of the groups, conditions, and in those interactions. 
The two-way mixed-design ANOVA for mean MSI predicted using the 6 DoF SVC-VV model revealed a significant main effect of conditions, with the LAD condition showing lower MSI than the WAD condition ($p=0.005$); however, no significant effect was found in the groups and their interaction.

Using the maximum MSI as an evaluation index, Figs. ~\ref{fig:MSI_resuts}~\subref{fig:MSI_mean_max_c} and~\subref{fig:MSI_mean_max_d} show the mean MSI predicted by the 6 DoF SVC and 6 DoF SVC-VV models, respectively.

As presented in Table~\ref{tab:ANOVA_Max_MSI}, a two-way mixed-design ANOVA for the maximum MSI predicted by the 6 DoF SVC model revealed no significant effect of the groups, conditions, and interactions. 
For the maximum MSI predicted using the 6 DoF SVC-VV model, the two-way mixed-design ANOVA revealed a significant main effect of conditions, with the LAD condition showing significantly lower MSI than the WAD condition ($p=0.003$); however, no significant effect was found in the groups and their interaction.

\subsection{Comparison between predicted MSI and reported MISC}

As shown in Fig.~\ref{fig:MISC_TS_all}, there were six participants (\#02, \#06, \#15, \#18, \#22, \#26) who did not get any motion sickness symptoms, \ie all MISC reported were zero, in both LAD and WAD conditions.
Considering that our proposed 6 DoF SVC-VV model was used to predict MSI, \ie, the percentage of participants who may vomit during the experiment, the predicted MSI was difficult to represent the individual features of participants who did not suffer from motion sickness.
Therefore, in the analysis in this subsection, we excluded data from these six participants.

\begin{figure}[!hb] 
\vspace{3mm}
\centering 
     \begin{subfigure}[b]{0.245\textwidth}
         \centering
         \includegraphics[width=\textwidth]{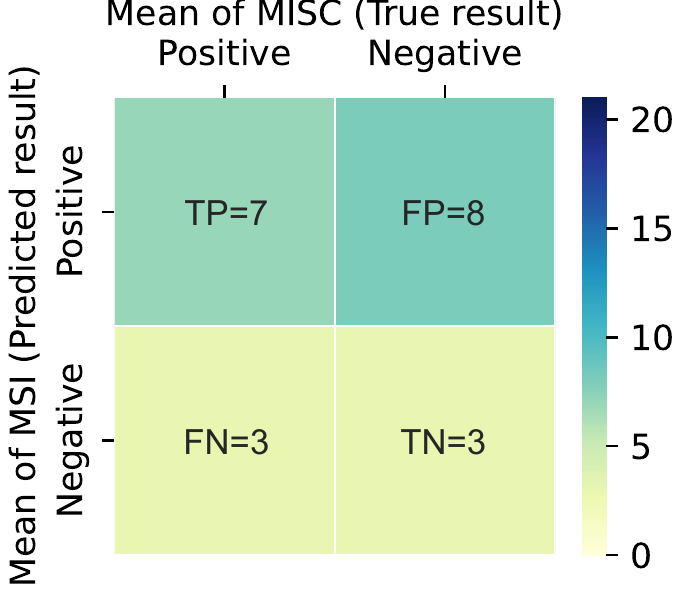}
         \caption{Mean of MSI by\\6 DoF SVC model}
         \label{fig:TP_FP_a}
     \end{subfigure}
     \hfill
     \begin{subfigure}[b]{0.245\textwidth}
         \centering
         \includegraphics[width=\textwidth]{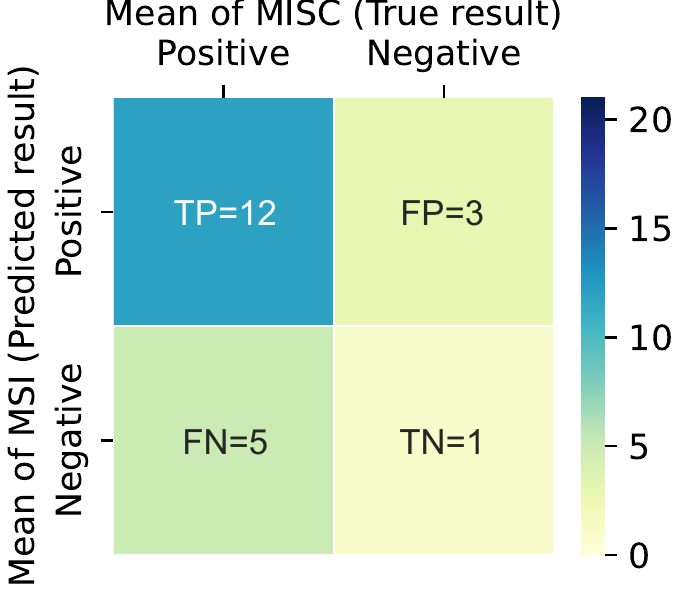}
         \caption{Mean of MSI by\\6 DoF SVC-VV model}
         \label{fig:TP_FP_b}
     \end{subfigure}
     \hfill
     \begin{subfigure}[b]{0.245\textwidth}
         \centering
         \includegraphics[width=\textwidth]{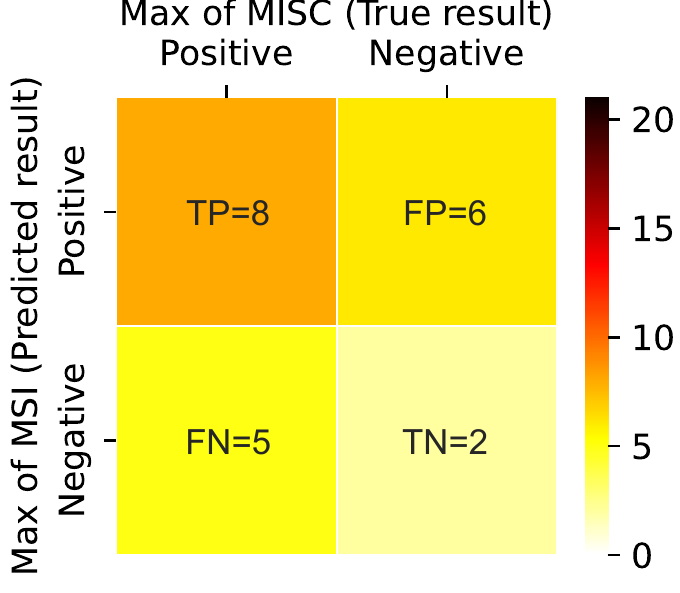}
         \caption{Maximum of MSI by\\6 DoF SVC model}
         \label{fig:TP_FP_c}
     \end{subfigure}
       \hfill
     \begin{subfigure}[b]{0.245\linewidth}
         \centering
         \includegraphics[width=\linewidth]{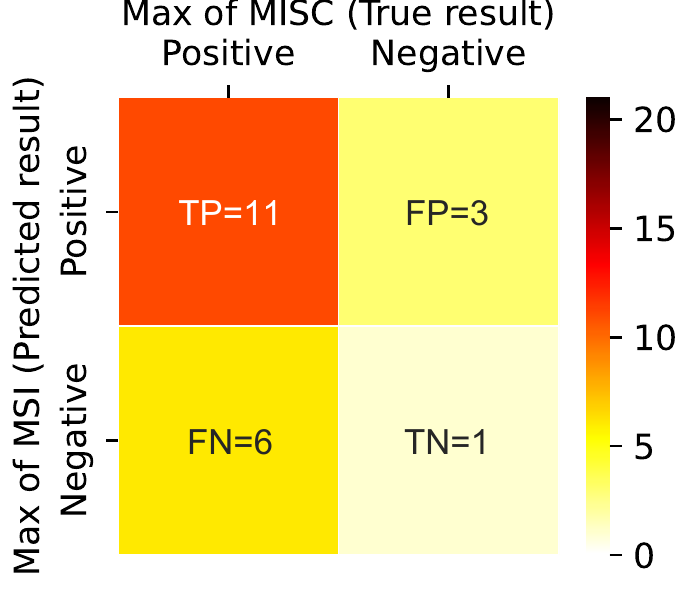}
         \caption{Maximum of MSI by\\6 DoF SVC-VV model}
         \label{fig:TP_FP_d}
     \end{subfigure}
\caption{Confusion matrices for the mean and maximum MSI predicted by 6 DoF SVC and 6 DoF SVC-VV models based on the mean and maximum MISC reported from 21 participants who felt symptoms of motion sickness in at least one of the riding conditions. As the true result, the mean of MISC had 15 positive cases and 6 negative cases; the Maximum of MISC had 14 positive cases and 7 negative cases. } 
\label{fig:TP_FP}
\end{figure}

By analyzing the MISC of the remaining 21 participants, the mean MISC had 15 positive cases and 6 negative cases; the maximum MISC had 14 positive cases and 7 negative cases. 
Taking the MISC results as the true result, confusion matrices of the mean and maximum MSI predicted by the 6 DoF SVC and 6 DoF SVC-VV models are shown in Fig.~\ref{fig:TP_FP}.
Regardless of whether for the mean or maximum MSI, the TP when using the 6 DoF SVC-VV model was higher than that when using the 6 DoF SVC-VV; however, the TN when using the 6 DoF SVC-VV model was lower than that when using the 6 DoF SVC model.

\begin{figure}[!hb] 
\centering 

     \begin{subfigure}[b]{0.495\textwidth}
         \centering
         \includegraphics[width=\textwidth]{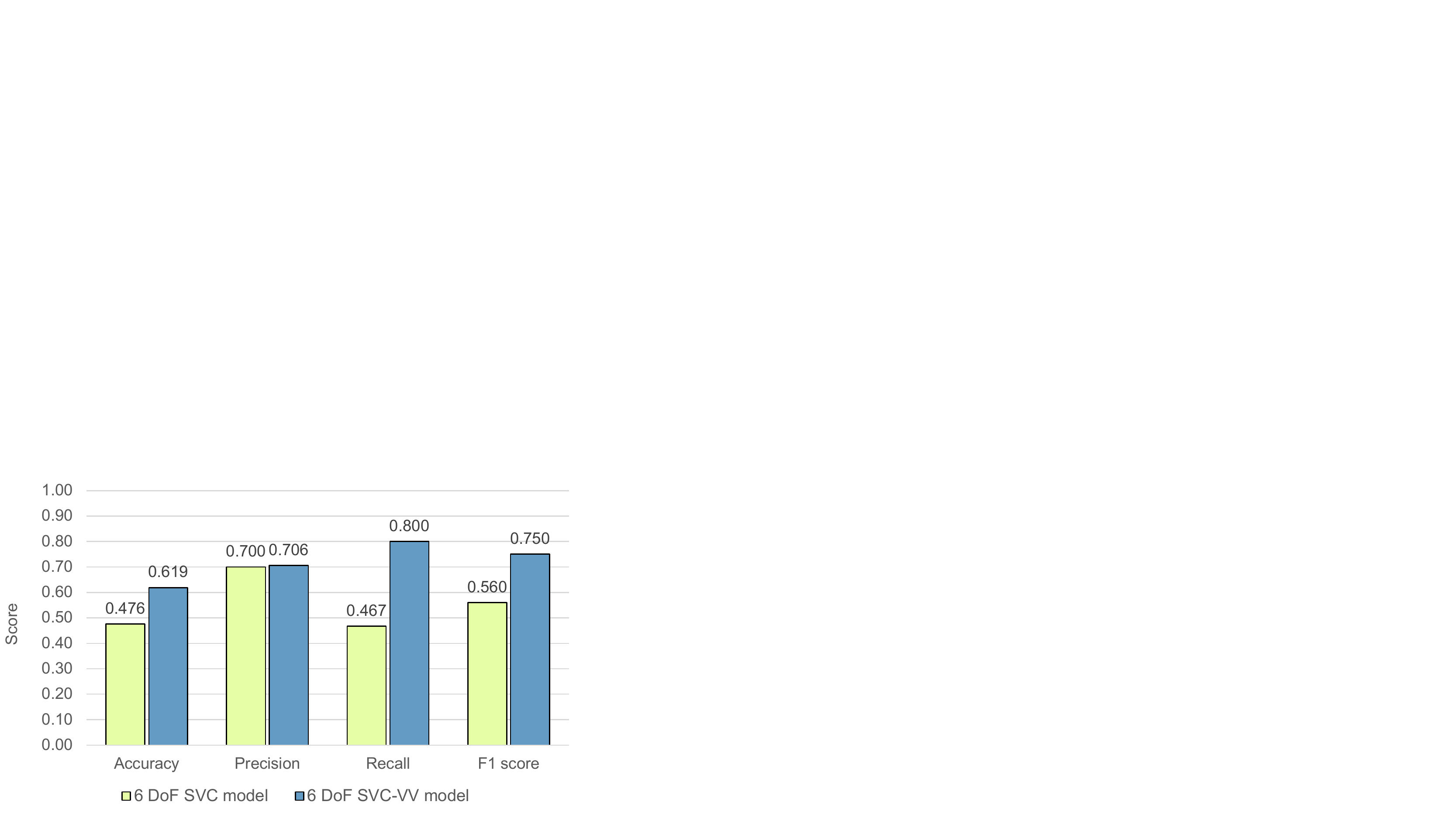}
         \caption{Mean of MSI}
         \label{fig:F1_score_a}
     \end{subfigure}
     \hfill
     \begin{subfigure}[b]{0.495\textwidth}
         \centering
         \includegraphics[width=\textwidth]{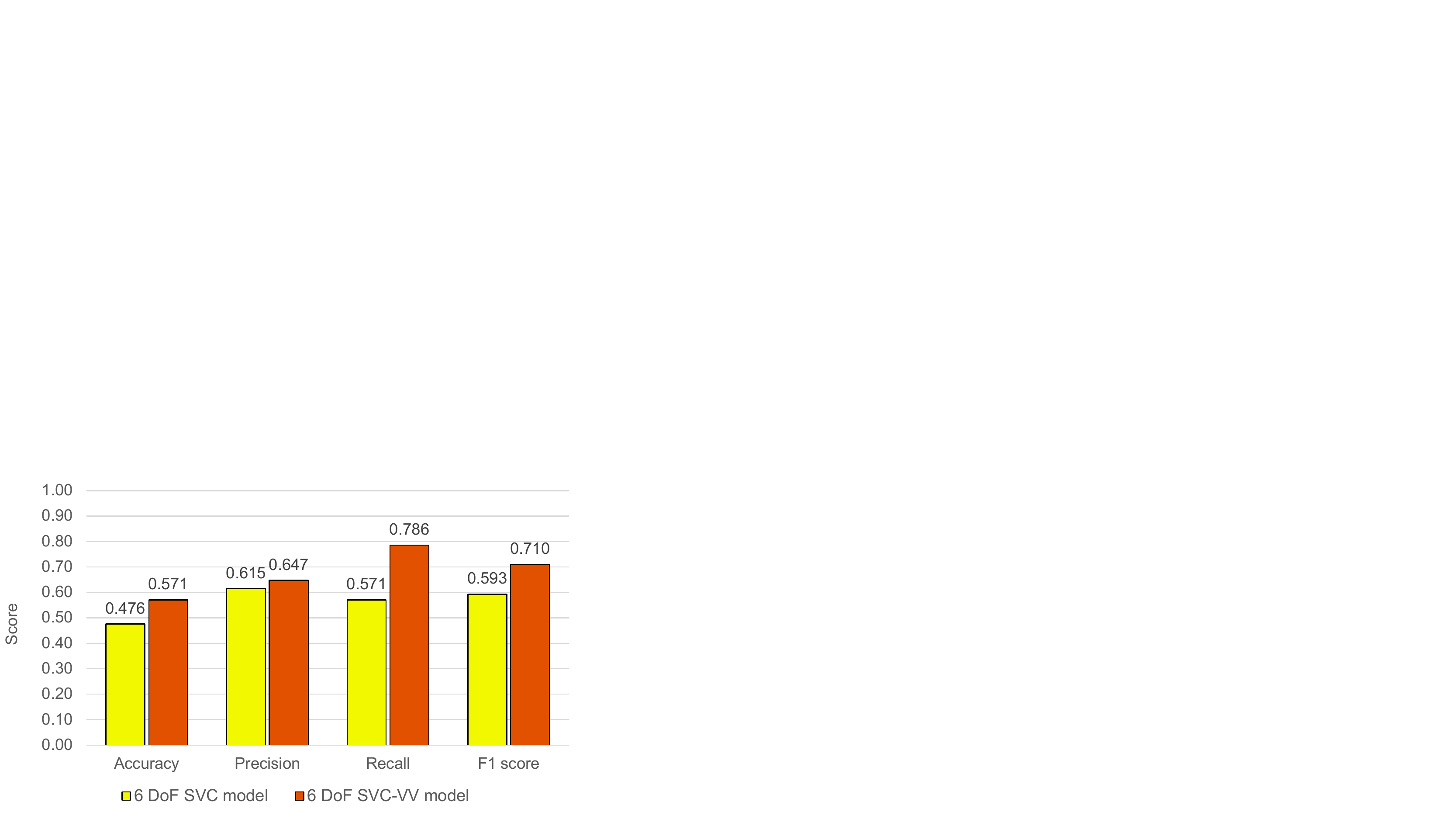}
         \caption{Maximum of MSI}
         \label{fig:F1_score_b}
     \end{subfigure}
\caption{Accuracy, precision, recall, and F1 score for the mean and maximum MSI predicted by the 6 DoF SVC and 6 DoF SVC-VV models based on the mean and maximum MISC reported from 21 participants who felt symptoms of motion sickness in at least one of the riding conditions.} 
\label{fig:F1_score}
\end{figure}

Based on these confusion matrices, the scores of accuracy, precision, recall, and F1-score for the mean and maximum MSI predicted by the 6 DoF SVC and 6 DoF SVC-VV models are shown in Fig.~\ref{fig:F1_score}.
Both the mean and maximum MSI predicted using the 6 DoF SVC-VV model had higher scores for accuracy, precision, recall, and F1 score than those predicted using the 6 DoF SVC model.
However, the difference between the precision scores of the mean and maximum MSI when these two models were used was small.

\section{DISCUSSIONS}

\subsection{Reported MISC}

The MISC reported by 27 participants at each minute is shown in Fig.~\ref{fig:MISC_TS_all}. 
We found that 21 of the 27 participants experienced motion sickness symptoms during the APMV riding experiment. 
This result illustrates that although APMVs are driven at a lower speed than vehicles, their passengers also have a high risk of motion sickness when it frequently avoids obstacles or other pedestrians. 
However, many other studies have reported that passengers are prone to motion sickness when using cars~\citep{turner1999motion, Wada2012, brietzke2021influence}.
or other vehicles, such as vessels~\citep{pepper1979repeated, turan2009motion}, and aircrafts~\citep{samuel2015airsickness}, we reported unprecedented results of motion sickness in passengers when using miniaturized autonomous vehicles, such as APMV.
Therefore, this study presents a novel issue regarding motion sickness and a new mindset for improving passenger ride comfort for researchers and manufacturers of miniaturized autonomous vehicles. 

As presented in Table~\ref{tab:MISC_ANOVA}, we did not find a significant effect of the experimental order of LAD and WAD on the MISC results based on the between-group design. 
For the within-group design, the results of the statistical tests validated our hypothesis H~1 in participants working with a tablet device on the APMV (WAD) produced significantly more profound motion sickness symptoms than when looking ahead (LAD) (Fig.~\ref{fig:MISC_mean_max} and Table~\ref{tab:MISC_ANOVA}). 
This result is consistent with the conclusions of ~\citep{ griffin2004visual,kato2006study,KARJANTO2018678, karjanto2018effect,kuiper2018looking,suwareducing,irmak2021objective}, where motion sickness is induced by visual obstruction when passengers ride in a car.

Furthermore, we found that, in Fig.~\ref{fig:MISC_mean_max}, the degree of motion sickness symptoms differed among participants. 
Such individual differences in motion sickness susceptibility were also reported in ~\citep{GOLDING2006237,irmak2021objective}. 
Although this study aimed to model motion sickness to predict MSI, considering these individual differences in the 6 DoF SVC-VV model is challenging future work.

\subsection{Calculated visual vertical}
The calculated visual vertical directions $\theta^{vv}$ by the proposed method were compared with the gravitational acceleration directions $\theta^{g}$ which were estimated from the measured acceleration ($\bm{f}=\bm{g}+\bm{a}$) by IMU attached to the participant’s head. 
The results in Fig.~\ref{fig:Corr_LAD_WAD} show that the mean of the correlation coefficients under the LAD condition was significantly higher than their correlation coefficients under the WAD condition, suggesting that the proposed visual vertical prediction method can: 1) calculate the visual vertical direction from environmental images with correlation to the direction of gravitational acceleration in the absence of visual occlusion; 2) represent the effect of visual occlusion on the prediction of visual vertical information from environmental images, \eg participants look-ahead and look-at-the-tablet device in an indoor environment. 

This conclusion is consistent with the results of our previous study~\citep{LIU2022_IV}, which focused on the use of APMV under outdoor conditions. 
However, the correlation coefficients between the calculated visual vertical direction and gravitational acceleration direction in this experiment were not as high as those in our previous study.
Two reasons can be considered:1) The experimental scenes in this study were indoors; thus, the contours of the objects affected the visual vertical prediction, \eg, tables and chairs placed in different directions.  
In \citep{LIU2022_IV}, the horizon line and contours of buildings in an open scene contribute to the prediction of the visual vertical direction.  
2) There was noise in estimating gravitational acceleration owing to the movement of APMV. 
In \citep{LIU2022_IV}, the calculated visual vertical directions were evaluated with the direction of acceleration $\bm{f}$ under a static condition (APMV was stopped, $\bm{a}=[0,0,0]^T$), in which measured acceleration $\bm{f}=\bm{g}$, whereas the present study used an acceleration signal, which is composed of the summation of inertial acceleration and gravitational acceleration, \ie $\bm{f}=\bm{g}+\bm{a}$.

\subsection{Predicted MSI}

The conventional 6 DoF SVC and 6 DoF SVC-VV models were 
used to predict MSI from IMU and camera data from 27 participants (Fig.~\ref{fig:MSI_resuts}).

As previously explained, MSI and MISC are different indicators for evaluating motion sickness, \ie MSI indicates the percentage of participants experiencing vomiting when exposed to motion for a certain time; MISC indicates each participant's subjective assessment of the severity of motion sickness. 
Compared to the MISC reported in Fig.~\ref{fig:MISC_TS_all}, the predicted MSI cannot represent the individual differences in motion sickness susceptibility.  

In a between-group design, 
the results in Tables~\ref{tab:ANOVA_Mean_MSI} and~\ref{tab:ANOVA_Max_MSI} show that the mean and maximum of the predicted MSI using the two models were not significantly different between the groups, that is, the order of LAD and WAD, is consistent with the MISC results (see Table~\ref{tab:MISC_ANOVA}). 

In a within-group design, 
the mean and maximum of the predicted MSI using the 6 DoF SVC model showed no significant difference between the LAD and WAD conditions.
This is inconsistent with the MISC results because the MISC in WAD is significantly higher than that in LAD (see Table~\ref{tab:MISC_ANOVA}).
Contrarily, the proposed 6 DoF SVC-VV model predicted a significantly higher MSI under WAD than under LAD with the same trend as the MISC reported by the participants (see Fig.~\ref{fig:MISC_mean_max} and Table~\ref{tab:MISC_ANOVA}). 
This implies that adding a visual vertical part to the conventional vestibular motion sickness 6DOF-SVC model facilitates the description of the difference in motion sickness under different visual conditions. 
 
In summary, the results obtained in this study imply that the proposed 6 Dof SVC-VV model can describe the difference in the severity of motion sickness for different vertical visual conditions, such as increased motion sickness when reading books during APMV while the conventional 6~Dof SVC model does not.

\subsection{Comparison between the predicted MSI and the reported MISC}

In this subsection, we discuss the performance of the predicted MSI by comparing it with the reported MISC. 

Figure~\ref{fig:TP_FP} shows the confusion matrices for the mean and Maximum of MSI predicted by the 6 DoF SVC and the 6 DoF SVC-VV models based on the mean and Maximum of MISC reported by 21 participants (excluding six participants who did not have any motion sickness symptoms).
The mean of MSI ( Fig.~\ref{fig:TP_FP}~(a) and (b)), when using 6~Dof SVC-VV model, TP=12 was higher than when using 6~Dof SVC model, \ie TP=7.
Similarly, the TP cases of the maximum MSI using the 6~Dof SVC-VV model was 11, which was also higher than the TP=8 obtained using the 6~Dof SVC model.
However, for both the mean and maximum of MSI, the number of correctly predicted negative cases (TN) by the 6~Dof SVC-VV model was smaller than when using the 6~Dof SVC model.
This may be caused by an excessive feedback gain $K_{vvc}$ of the visual vertical conflict.
We will address this issue in future studies by adjusting the balance of parameters $K_{vc}$ and $K_{vvc}$.

Based on the confusion matrices above, the accuracy, precision, recall, and F1 score for the mean and Maximum of MSI predicted by the 6 DoF SVC and 6 DoF SVC-VV models are shown in Fig.~\ref{fig:F1_score}.
The accuracy and F1 score are the overall evaluations of the prediction results.
The accuracy and F1 scores of the 6 DoF SVC-VV model were higher than those of the 6 DoF SVC model for both the mean and maximum values of the MSI.
Moreover, precision and recall are sub-scores of the F1 score.
For both the mean and maximum of the MSI, the precision scores of the 6 DoF SVC-VV and 6 DoF SVC models were similar; however, the recall scores of the 6 DoF SVC-VV model were higher than those of the 6 DoF SVC model.
Based on Table~\ref{tab:evaluation_index}, both models had the same performance in predicting the correct positive results over all positive predictions; however, the 6 DoF SVC-VV model had a better performance in predicting correct positive results over all positive true results.

\subsection{Limitations}

The visual vertical prediction method proposed in this study can only predict the visual vertical direction in a 2D plane.
Therefore, the visual vertical direction changes caused by head rotations along the pitch axis cannot yet be calculated.

The parameters in Table~\ref{tab:parameter} were obtained from the conventional 6~DoF SVC model of a previous study~\citep{Inoue2022}, which did not include visual-vestibular interaction.
Therefore, these parameters may not be optimal for the proposed 6~DoF~SVC-VV model.

All participants were in their 20s. 
A broader demographic survey is necessary, particularly for the elderly, who are potential wheelchair users.

Furthermore, the proposed 6~DoF~SVC-VV model cannot be used to represent the individual traits of motion sickness susceptibility because this model is designed to predict MSI.
Particularly, it is difficult to apply this model to people who are extremely insensitive to motion sickness.

\subsection{Future works}

The visual vertical prediction method was improved to extract 3D visual vertical features from image data. 
This will further help the 6 DoF SVC-VV model represent the change in the visual vertical direction owing to head rotations on the pitch axis.

Although the 6 DoF SVC-VV model uses the parameters optimized by~\citet{Inoue2022}, a new parameter $K_{vvc}$, which is the feedback gain of the visual vertical conflict, has not yet been optimized. 
We will address this issue in future studies by adjusting the balance of parameters $K_{vc}$ and $K_{vvc}$.

Furthermore, because the proposed 6~DoF~SVC-VV model cannot be used to represent the individual traits of motion sickness susceptibility, we will develop a new MISC prediction model based on the 6~DoF~SVC-VV model based on the model proposed in~\citep{irmak2021objective}.

Moreover, we consider that working with a tablet device during riding APMV hinders the visual vertical perception of passengers and their motion perception through dynamic vision. 
Therefore, the integration of the proposed 6~DoF~SVC-VV model with the visual flow, referring to \citet{wada2020computational}, is an important future direction.

\section{CONCLUSION}
To model motion sickness in passengers under different visual conditions while using the APMV, this study proposes a new computational model of SVC theory for predicting motion sickness that considers the interactions between vertical perception from the visual and vestibular systems. 
We added a module for visual vertical perception to the 6~DoF SVC model in ~\citep{Inoue2022}. 
Therefore, we proposed a visual vertical prediction method based on an image processing technique.

In the experiment, 27 participants experienced APMV with two visual conditions: looking ahead (LAD) and working with a tablet device (WAD). 
Of these, 21 participants reported motion sickness symptoms, particularly in the WAD condition.
Furthermore, based on the MISC reported by the participants, we found that the proposed 6~DoF~SVC-VV model more accurately predicted MSI than the conventional 6~DoF SVC model without visual input when the visual vertical direction and direction of gravitational acceleration differed, such as when participants worked with a tablet device while using an APMV.

\section*{ACKNOWLEDGMENTS}
This work was supported by JSPS KAKENHI Grant Numbers 21K18308 and 21H01296, Japan. 

\section*{CRediT authorship contribution statement}

\textbf{Hailong Liu}: Conceptualization, Investigation, Resources, Methodology, Validation, Formal analysis, Visualization, Writing - Original Draft, Writing - Review \& Editing.

\textbf{Shota Inoue}: Conceptualization, Methodology, Writing - Review \& Editing.

\textbf{Takahiro Wada}: Conceptualization, Methodology, Writing - Review \& Editing, Project administration, Funding acquisition.

\section*{Declaration of Competing Interest}

The authors declare that they have no known competing financial interests or personal relationships that could have appeared to influence the work reported in this paper.

\bibliographystyle{cas-model2-names} 
\bibliography{sample.bib}

\end{document}